\documentclass[fleqn,usenatbib]{mnras}

\usepackage{newtxtext,newtxmath}
\usepackage[T1]{fontenc}

\DeclareRobustCommand{\VAN}[3]{#2}
\let\VANthebibliography\thebibliography
\def\thebibliography{\DeclareRobustCommand{\VAN}[3]{##3}\VANthebibliography}

\usepackage{graphicx}	% Including figure files
\usepackage{amsmath}	% Advanced maths commands
\graphicspath{{images/}}
\newcommand{\kms}{\,km\,s$^{-1}$}

\title[Galactic tides and the Crater II dwarf spheroidal]{Galactic tides and the Crater II dwarf spheroidal: a challenge to LCDM?} 

\author[A. Borukhovetskaya et al.]{
Alexandra Borukhovetskaya,$^{1}$\thanks{E-mail: asya@uvic.ca}
Julio F. Navarro$^{1}$,
Rapha\"{e}l Errani$^{1,2}$,
and Azadeh Fattahi$^{3}$
\\
$^{1}$Department of Physics and Astronomy, University of Victoria, Victoria, BC V8P 5C2, Canada\\
$^{2}$ Universit\'e de Strasbourg, CNRS, Observatoire astronomique de Strasbourg, UMR 7550, F-67000 Strasbourg, France\\
$^{3}$ Institute for Computational Cosmology, Department of Physics, University of Durham, South Road, Durham DH1 3LE, UK
}

\date{Accepted 2022 March 2. Received 2022 March 1; in original form 2021 December 2}

\pubyear{2021}

\begin{document}
\label{firstpage}
\pagerange{\pageref{firstpage}--\pageref{lastpage}}
\maketitle

\begin{abstract}
  The unusually low velocity dispersion and large size of Crater II pose a challenge to our understanding of dwarf galaxies in the Lambda Cold Dark Matter (LCDM) cosmogony. The low velocity dispersion suggests either a dark halo mass much lower than the minimum expected from hydrogen cooling limit arguments, or one that is in the late stages of extreme tidal stripping. The tidal interpretation has been favoured in recent work and is supported by the small pericentric distances consistent with available kinematic estimates. We use \textit{N}-body simulations to examine this interpretation in detail, assuming a Navarro-Frenk-White (NFW) profile for Crater II's progenitor halo. Our main finding is that, although the low velocity dispersion can indeed result from the effect of tides, the large size of Crater II is inconsistent with this hypothesis. This is because galaxies stripped to match the observed velocity dispersion are also reduced to sizes much smaller than the observed half-light radius of Crater II. Unless its size has been substantially overestimated, reconciling this system with LCDM requires that either (i) it is not bound and near equilibrium (unlikely, given its crossing time is shorter than the time elapsed since pericentre), or that (ii) its progenitor halo deviates from the assumed NFW profile. The latter alternative may signal that baryons can affect the inner halo cusp even in extremely faint dwarfs or, more intriguingly, may signal effects associated with the intimate nature of the dark matter, such as finite self-interactions, or other such deviations from the canonical LCDM paradigm.
\end{abstract}


%%%%%%%%%%%%%%%%%%%%%%%%%%%%%%%%%%%%%%%%%%%%%%%%%%

%%%%%%%%%%%%%%%%% BODY OF PAPER %%%%%%%%%%%%%%%%%%

\section{Introduction}
\label{SecIntro}
The Crater II dwarf spheroidal (dSph) is a distant Milky Way (MW) satellite discovered in imaging data from the ATLAS survey at the VLT Survey Telescope  by \citet{Torrealba2016}. Its unusual properties were immediately apparent. Although its total luminosity ($M_V\sim -8$) is comparable to that of the faintest ``classical'' dSphs, such as Draco and Ursa Minor, Crater II's enormous size (projected half-light radius, $R_{1/2}\sim 1$ kpc, and spanning nearly $5^\circ$ across the sky) is comparable to that of Fornax, a dSph more than $100$ times more luminous. Indeed, Crater II is one of the lowest surface brightness galaxies ever discovered, several decades fainter than the ultra-diffuse galaxy population identified by surveys such as Dragonfly \citep{vanDokkum2015}.

These unusual photometric properties are compounded by equally unusual kinematics. The most recent estimates put the velocity dispersion of its stars at $\sigma_{\rm los}\approx 2.3$\kms, one of the lowest amongst all known dSphs but still large enough to imply the presence of large amounts of dark matter ($M/L\approx 50$: \citet{Caldwell2017,Fu2019}; $M/L\approx 30$: \citet{Ji2021}). Such uncommon properties point to atypical formation paths. Suggestions include the possibility that Crater II formed in a halo of unusually low density \citep{Amorisco2019}, or that its structure was severely affected by Galactic tides after formation, or both.

The latter explanation is currently the most popular. \citet{Frings2017} and \citet{Applebaum2021}, for example, report simulations of satellites whose properties are similar to those of Crater II after undergoing severe tidal stripping. Similar conclusions were reported by \citet{Sanders2018}, who cautioned, however, that it was difficult to account simultaneously for the velocity dispersion and size of Crater II in cuspy dark matter halos such as those expected in the Lambda Cold Dark Matter (LCDM) cosmogony \citep{Navarro1996a,Navarro1997}. \citet{Fattahi2018} extrapolated the ``tidal tracks'' of \citet{penarrubia2008} and \citet{EPT15} to reach a similar conclusion, noting that explaining Crater II through tidal stripping implied the rather extreme case of a progenitor which had lost more than $99\%$ of its initial stellar content to tides.

Despite these difficulties, consensus for a tidal interpretation of the unusual properties of Crater II \citep[and for those of Antlia II, another ``feeble giant'' MW satellite;][]{Torrealba2019} seems to have emerged. In the case of Antlia II, the case for tides has been strengthened by the detection of a clear velocity gradient aligned with the orbital path indicated by the latest proper motions from Gaia \citep{Ji2021}.

No such gradient, however, is clearly present in Crater II, at least over the area surveyed spectroscopically. This is perhaps not surprising: Crater II's negative Galactocentric radial velocity suggests that it is at present just past apocentre; together with its large distance ($\sim 120$ kpc), this implies that its most recent pericentric passage must have occurred about $\sim 1$\,Gyr ago. As discussed\footnote{See also earlier work by \citet{Aguilar1986} and \citet{Navarro1990}.} by \citet{Penarrubia2009}, the inner regions of a bound tidal remnant relax quickly, pushing all signatures of  tidal disturbance to regions where the local crossing time exceeds the time elapsed since pericentre, $t-t_p$ (see their Eq.~5). Assuming $(t-t_p) \sim 1$\,Gyr, and using the measured velocity dispersion, this implies that the inner $R \lesssim 1.3$\,kpc  should be close to dynamical equilibrium. This region is similar to the area over which  radial velocities are available \citep[$\sim 0.65$ deg from the dwarf's centre;][]{Ji2021}.

This insight simplifies considerably the analysis, as it disfavours interpretations where the singular properties of Crater II are due to large, transient departures from equilibrium. If Galactic tides have indeed been responsible for shaping Crater II, then its properties must be consistent with the structure of the remnants of tidally stripped LCDM subhalos, an issue that has been studied extensively over the years using \textit{N}-body simulations \citep[see; e.g.,][]{Hayashi2003,penarrubia2008,EPT15}. The latest work suggests that LCDM subhalos, if well approximated by cuspy Navarro-Frenk-White \citep[hereafter NFW, see][]{Navarro1996a,Navarro1997} profiles, should almost always leave behind a bound remnant \citep{Penarrubia2010,vdBOgiya2018, vdb2018}. The remnant properties are fully specified by the initial characteristic crossing time of the subhalo, by the number of orbits completed, and by the orbital time at pericentre \citep{EN21}.

This discussion thus calls into question whether Crater II is actually a stellar system inhabiting a subhalo nearing full tidal disruption.  Are the peculiar size and kinematics of Crater II actually consistent with the tidal remnants of LCDM halos?

We address this issue here using \textit{N}-body simulations of the tidal evolution of NFW subhalos in orbits chosen to match the observed present-day position and velocity of Crater II in the MW halo potential. We choose cuspy NFW halos for this work not only because this is one of the best tested LCDM predictions, but also because the halo inner structure of galaxies as faint as Crater II should be relatively unaffected by the effects of baryons, which could in principle alter the inner density cusp in more massive galaxies \citep[see; e.g.,][and references therein]{Navarro1996b, Read2005, Pontzen2012,DiCintio2014}.

This paper is organized as follows. We first summarize the observed properties of Crater II, as well as our numerical setup,  in Sec.~\ref{sec:num_setup}. The results of the simulations, for both dark matter and stellar tracers, are analysed and presented in Sec.~\ref{sec:results}. We end with a brief summary and a discussion of our results in a cosmological context in Sec.~\ref{SecConc}. 

%%%%%%

\section{Observations and Simulations}
\label{sec:num_setup}
This section summarises the observed properties of Crater II, and outlines the numerical setup of the \textit{N}-body simulations used to model the tidal evolution of Crater II in the gravitational potential of the Milky Way.

\subsection{Observed properties of Crater II}
\label{sec:observations}
\begin{figure*}
\includegraphics[width=\textwidth]{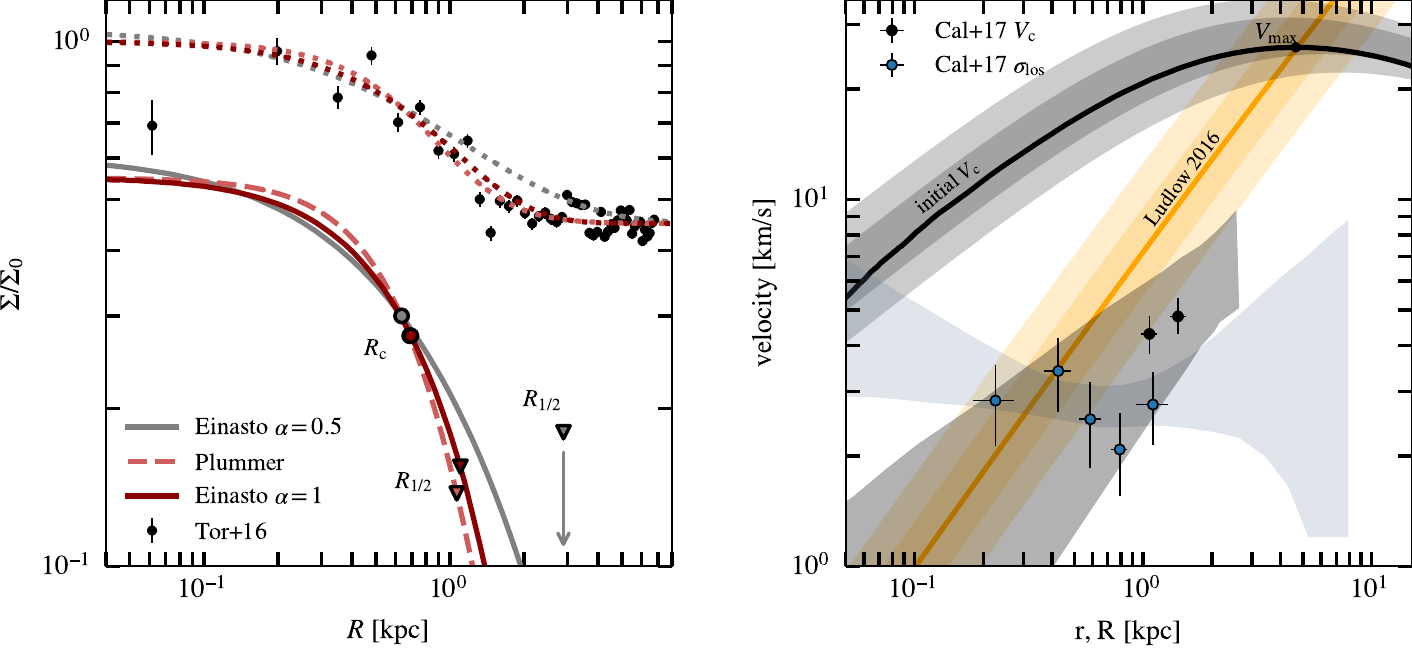}
 \caption{Observed structural parameters of Crater II. The left panel shows the observed surface brightness taken from \citet{Torrealba2016} together with  fits adopting different models for the density profile of the stellar component. Light red curves correspond to a \citet{Plummer1911} fit with the parameters of \citet{Torrealba2016}. Dotted profiles include the contribution of a constant background level equal to $0.45\times$ the central surface brightness, $\Sigma_0$. Dark red curves correspond to an $\alpha=1$ Einasto profile (Eq.~\ref{eq:Einasto}), which fits the Plummer projected density profile quite well. Grey curves show an $\alpha=0.5$ Einasto profile with similar core radius, $R_{\rm c}$ (circles), but a much larger half-light radius ($R_{1/2}$, triangles) than the other profiles. All seem to provide acceptable fits to the data, implying that, unlike $R_{\rm c}$, Crater II's $R_{1/2}$ is not well constrained.
   The right panel shows the measured velocity dispersion of Crater II (blue circles, and shaded area) as well as corresponding estimates for the circular velocity , taken from \citet{Caldwell2017}. The  black line (and shaded area) shows the circular velocity profile of our Crater II NFW model before stripping, motivated by results from the APOSTLE simulation for galaxies of similar stellar mass \citep{Fattahi2018}. The yellow line shows the $r_{\rm max}$-$V_{\rm max}$ relation for LCDM halos at $z=0$ from  \citet{Ludlow2016}. The inner shaded bands correspond to $\pm 1$-sigma scatter in concentration and the outer, fainter bands to $\pm 2$-sigma.}
 \label{fig:data}
\end{figure*}

The photometric properties of Crater II were estimated by \citet{Torrealba2016} in its discovery paper. These authors estimate a projected half-light radius of order $R_{1/2}\sim 1.066$\,kpc by fitting a Plummer model to star counts.  We reproduce their published data in the left panel of Fig.~\ref{fig:data}, which shows that the number density profile spans only a factor of $\sim 2$ in density before becoming dominated by foreground stars. This results in large uncertainties in estimates of the half-light radius, depending on the assumed profile shape. This is illustrated by the grey and dark red lines in Fig.~\ref{fig:data}, which both fit the observations fairly well although their half-light radii (as marked by triangles) differ by a factor of $\sim 3$. A much better constrained parameter is the core radius, $R_{c}\approx 700$ pc, defined as the radius where the surface brightness of a galaxy drops by a factor of two from the central value.

Stellar half-light radii are also difficult to measure in numerical simulations, particularly in cases where tides have led to substantial amounts of mass loss. In these cases, estimated $R_{1/2}$ values depend critically on which stars are included in the analysis. Including weakly bound, or escaping, stars, for example, typically results in poorly-defined estimates vulnerable to sizeable transient fluctuations.  Because of this, we shall adopt $R_{c}$ (marked with circles in the left panel of Fig.~\ref{fig:data}) rather than $R_{1/2}$ (triangles) as the characteristic photometric radius of Crater II, although our main conclusions  do not depend critically on this choice. For a Plummer model, which is often adopted in observational studies to fit the stellar density profile, $R_{\rm c}\approx0.6\, R_{1/2}$.

The right-hand panel of Fig.~\ref{fig:data} summarizes some of the kinematic information available for Crater II, compiled from \citet{Caldwell2017}. (We shall hereafter use lowercase $r$ to denote 3D radii and uppercase $R$ for projecte radii.) The observed velocity dispersion profile is shown in grey-blue. The black circles indicate corresponding circular velocity estimates obtained using the  \citet{Wolf2010} and \citet{Walker2009} mass estimators \footnote{These mass estimates are supported by the full Jeans modelling of \citet{Caldwell2017} (shaded region).}.
 
For comparison, we show with a thick black curve the (NFW) circular velocity profile expected for an {\it isolated} dwarf galaxy of stellar mass comparable to that of Crater II (assumed to be $M_\star\approx 2.56\times10^5$\,M$_\odot$,  see Sec.~\ref{sec:cra2_model} for details). According to the results of the  APOSTLE cosmological hydrodynamical simulations \citep{Fattahi2018}, galaxies like Crater II are expected to form in halos with peak circular velocity $V_{\rm max}\sim 26$\kms, or a virial\footnote{We define virial quantities as those within a sphere of mean density equal to 200 times the critical density for closure, $\rho_{\rm crit}=3H_0^2/8\pi G$. We use the subscript ``200'' to indicate virial quantities.}  mass $M_{200}\sim 2.7\times10^{9}$\,M$_\odot$. For such model to be viable, it is clear from Fig.~\ref{fig:data}  that Galactic tides must have led to a large depletion of the original dark matter content of Crater II.

\subsection{Milky Way potential model}
\label{sec:host}
\begin{table*}
	\centering
	\caption{Parameters of the analytical, static Milky Way potential used in this study. The model is a re-parametrisation of the \citet{McMillan2011} model, as discussed in \citet{Errani2020}.}
	\label{tab:MW}
	\begin{tabular}{lllll}
		\hline
		Component & Functional form & & & \\
		\hline
		Disk (thin) & \citet{Miyamoto1975} & $M=5.9\times10^{10}\,\rm M _\odot$ & $a_{\rm d} = 3.9$\,kpc & $b_{\rm d} = 0.31$\,kpc \\
		Disk (thick) & \citet{Miyamoto1975} & $M=2.0\times10^{10}\,\rm M _\odot$ & $a_{\rm d} = 4.4$\,kpc & $b_{\rm d} = 0.92$\,kpc \\
		Bulge & \citet{Hernquist1990} & $M=2.1\times10^{10}\,\rm M _\odot$ & $a=1.3$\,kpc & \\
		DM Halo & \citet{Navarro1997} & $M_{200}=1.15\times10^{12}\,\rm M _\odot$ & $r_{\rm s} = 20.2$\,kpc & $c=r_{200}/r_{\rm s}=9.5$\\
		\hline
	\end{tabular}
      \end{table*}

The Milky Way host is represented by an analytical, static potential, which is comprised of an axisymmetric two-component \citet{Miyamoto1975} disk, a \citet{Hernquist1990} bulge, and an NFW dark matter halo. The model parameters are taken from \citet{Errani2020} -- i.e., chosen to approximate the \citet{McMillan2011} model, wherein at the solar circle $R_0 = 8.29$\,kpc, the circular velocity is $V_\mathrm{c} = 240$\,km\,s$^{-1}$. The thick and thin \citet{Miyamoto1975} disks are each parametrized by a disk mass $M$, a scale length $a_{\rm d}$, and a scale height $b_{\rm d}$. The \citet{Hernquist1990} bulge is similarly defined with a total mass $M$ and scale length $a$. Finally, the Milky Way dark matter halo can be characterized by an NFW halo with a scale radius $r_{s}$ and a virial mass, $M_{200}$. The host parameters  are summarized in Table~\ref{tab:MW}.

\subsection{Orbits}
\label{sec:orbits}
\begin{table*}
	\centering
	\caption{Current observational constraints, as well as parameters of the three orbits explored using $N$-body simulations in this study. The \citet{Fritz2018} proper motions are followed first by the statistical error and second by the systematic error. Orbits 1 and 2 are the orbits corresponding to the 16th and 84th percentile of the distribution in pericentres obtained with the proper motions of \citet{Kallivayalil2018}. Orbit 3 is the orbit corresponding to the median observed quantities of \citet{McConnachie2020-DR3}. Pericentres and apocentres are computed for the Milky Way potential model discussed in section \ref{sec:host}.}      
	\label{tab:CraterII_orbit}
	\begin{tabular}{lcccccc}
    \hline
		 \bf{observation}                & $\alpha$        & $\delta$           & distance          &     $\mu_{\alpha^*}$  & $\mu_\delta$        & $v_r$\\
                    		 &  &  & (kpc) & (mas\,yr$^{-1}$) & (mas\,yr$^{-1}$) & (km\,s$^{-1}$)\\ \hline
		 & $11^h 49^m 12^s~^{(1)}$ & $-18^\circ 24' 0''~^{(1)}$ & $117.5 \pm 1.1~^{(1)}$    &     $-0.246 \pm 0.052~^{(3)}$ &  $-0.227 \pm 0.026~^{(3)}$ & $87.5 \pm 0.4~^{(2)}$\\ 
		  &   &   &   &    $-0.184 \pm 0.061\pm 0.035~^{(4)}$ &  $-0.106 \pm 0.031\pm 0.035~^{(4)}$ &  \\ 
		  &   &   &   &    $-0.07 \pm 0.02~^{(5)}$ &  $-0.11 \pm 0.01~^{(5)}$ &  \\ \hline
	\multicolumn{7}{c}{	$^{(1)}$ \citet{Torrealba2016}, $^{(2)}$ \citet{Caldwell2017}, $^{(3)}$ \citet{Kallivayalil2018}, $^{(4)}$ \citet{Fritz2018}, $^{(5)}$ \citet{McConnachie2020-DR3} } \\[0.3cm]
		\hline
		\bf{model parameters} & pericentre & apocentre & distance & $\mu_{\alpha^*}$ & $\mu_\delta$ & $v_r$\\
		 & (kpc) & (kpc) & (kpc) & (mas\,yr$^{-1}$) & (mas\,yr$^{-1}$) & (km\,s$^{-1}$)\\
		\hline
    orbit 1 & 4.23 & 130       & 116 & -0.169 & -0.267 & 87.8\\
		{orbit 2} & 15.5 & 133    & 117 & -0.102 & -0.225 & 87.2\\
		orbit 3 & 37.4 & 139    & 118 & -0.07 & -0.11 & 87.5\\
		\hline

	\end{tabular}
 \end{table*}
 
The orbit of Crater II in the assumed Milky Way potential may be estimated from its present-day Galactocentric position and velocity, as inferred from its sky position, radial velocity, distance, and proper motion, assuming that the effects of dynamical friction can be neglected. Of the observed parameters  (summarised in Table~\ref{tab:CraterII_orbit}), the proper motions contribute the majority of the uncertainty budget.

\begin{figure}
	\includegraphics[width=1\columnwidth]{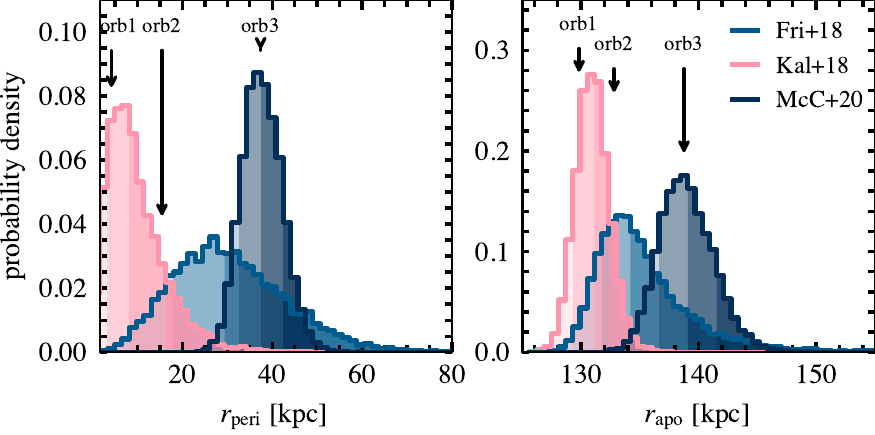}
        \caption{The probability distributions of Crater II pericentric (left panel) and apocentric (right panel) distances obtained from the proper motions, distance, and radial velocity, together with their reported uncertainties. Each colour corresponds to different proper motion estimates, from  \citet[][pink]{Kallivayalil2018}, \citet[][blue]{Fritz2018}, and \citet[][black]{McConnachie2020-DR3}. The three orbits selected to explore the available parameter space are shown with arrows pointing down. Regions beneath the histograms darken at the 16th, 50th, and 84th percentile. }
    \label{fig:histograms}
  \end{figure}

Assuming $\mu_{\alpha^*} = -0.246 \pm 0.052$ mas yr$^{-1}$, $\mu_\delta = -0.227 \pm 0.026$ mas yr$^{-1}$\citep{Kallivayalil2018}\footnote{Where with $\mu_{\alpha^*}$ we designate the proper motion in $\alpha$ including the $\cos(\delta)$ factor.}, for example, these uncertainties result in a fairly broad distribution of possible pericentric and apocentric distances, shown by the pink histograms in Fig.~\ref{fig:histograms}. Different estimates for the proper motion will, of course, yield different orbits. We show also in Fig.~\ref{fig:histograms} the pericentric and apocentric distributions corresponding to proper motion estimates from \citet{McConnachie2020-DR3} and \citet{Fritz2018}, illustrating that a wide range of orbits are permissible given the current data.

Because of the large allowed range, we have explored three different orbits, varying mainly the orbital pericentric distance (the most critical parameter for tidal effects), $r_\mathrm{peri} = 4.2$, $16$, and $37$\,kpc.  The corresponding orbits are hereafter referred to as orbits 1, 2 and 3, respectively, with parameters summarised in Table~\ref{tab:CraterII_orbit}. Initial conditions for the orbits of  $N$-body runs are obtained by integrating them  backwards in time for 10\,Gyr. The shape of the resulting orbits are shown in Fig.~\ref{fig:xyz} in a co-ordinate system where the Sun is located at $(X,Y,Z)_\odot=(-8.3,0,0)$\,kpc, and the velocity of the local standard of rest is in the positive $Y$ direction. 

\subsection{Crater II model}
\label{sec:cra2_model} 

\subsubsection{Dark matter component}
\begin{table}
\centering
\caption{Current properties of Crater II ($M_\star, V_{1/2}, r_{1/2}$) and inferred structural parameters at infall of the dark matter halo ($V_{\rm max}, r_{\rm max}$) and stellar component ($\alpha, R_\mathrm{core}, M_\star$). Half-light radius $r_{1/2}$ and circular velocity $V_{1/2}$ at the half-light radius are as in \citet{Caldwell2017}, taking $r_{1/2}=4/3R_{1/2}$.}
\label{tab:CraterII_structure}
\begin{tabular}{lccc}
		\hline
		\bf{observation}&$M_\star$ & $V_{1/2}$ & $r_{1/2}$\\
		                             &($10^5$\,M$_\odot$) & (km\,s$^{-1}$) & (kpc)\\
		\hline
		                             &$2.56$ & $4.8\substack{+0.6 \\ -0.5}$ & $1.421\pm 0.112$\\ \hline \\[0.2cm]
		\hline
		
		\bf{initial halo}& $V_\mathrm{max}$ & $r_\mathrm{max}$& \\
		                         & (km\,s$^{-1}$) & (kpc)& \\
		\hline		 
		                         & $25.9$ & 4.70\\ \hline \\[0.2cm]
		\hline 
		\bf{initial stars}& $\alpha$ & $R_\mathrm{core}$&  \\
		                         &   & (kpc)&  \\
		\hline		 
		E1& 1 & 0.46\\
		E2& 1 & 0.91\\
		E3& 1 & 1.79\\\hline
\end{tabular}
\end{table}

The Crater II halo is modelled as an equilibrium \textit{N}-body realisation of the NFW density profile,
\begin{equation}
    \rho_\mathrm{NFW}(r)=\frac{M_{200}}{4\pi r_s^3}\frac{(r/r_s)^{-1}(1+r/r_s)^{-2}}{[\ln(1+c)-c/(1+c)]}.
	\label{eq:NFW}
\end{equation}
This profile is fully specified by two parameters; e.g., a virial mass, $M_{200}$, and concentration, $c=r_{200}/r_s$, or, alternatively, by a maximum circular velocity, $V_\mathrm{max}$, and the radius at which it is reached, $r_\mathrm{max}$. 

Cosmological hydrodynamical simulations, such as the APOSTLE suite of Local Group simulations \citep{Sawala2016,Fattahi2016a}, have shown that $V_\mathrm{max}$ correlates strongly with galaxy stellar mass, $M_\star$ \citep[see also][for a recent compilation]{Santos-Santos2021}. We adopt a stellar mass for Crater II of $M_\star = 2.56 \times 10^5\,\mathrm{M_\odot}$, computed from the absolute magnitude of \citet{Torrealba2016} and assuming $M_\star/L_V=1.6$, typical of dwarf galaxies in the Local Group \citep{Woo2008}.

As in \citet{Borukhovetskaya2021}, we estimate $V_\mathrm{max}$ (before tidal effects) using the empirical fit\footnote{$M_\star = M_0 v^\alpha \exp(-v^\gamma)$, where $v=V_\mathrm{max}/50$\kms, and $(M_0, \alpha, \gamma)=(3.0\times10^8\, M_\odot, 3.36, -2.4)$.} to the $M_\star$-$V_\mathrm{max}$ correlation in APOSTLE from \citet{Fattahi2018}. The characteristic radius $r_\mathrm{max}$ then follows from the \citet{Ludlow2016} parametrisation of the LCDM halo mass-concentration relation at redshift $z=0$. As listed in Table~\ref{tab:CraterII_structure}, the resulting NFW profile has $V_\mathrm{max} = 25.9$\,km s$^{-1}$ and $r_\mathrm{max} = 4.7$\,kpc, or, in terms of virial mass and concentration, $M_{200} = 2.72\times10^{9}$\,M$_\odot$ and $c=13.6$. The corresponding circular velocity profile is shown in the right-hand panel of Fig.~\ref{fig:data}.

\subsubsection{Stellar component}

The stellar component of Crater II is modelled assuming that it contributes negligibly to the total gravitational potential. We consider, in particular, stellar components modelled as Einasto profiles \citep{Einasto1965}, \begin{equation}
    \rho_{\rm E}(r)=\rho_{\rm E0} \exp\left[-\left( r/r_{\rm E} \right)^{\alpha}\right],
	\label{eq:Einasto}
\end{equation}
with $\alpha=1$ (i.e., exponential spheres) and three different core radii: $R_{c}=0.5$, $0.9$, and $1.7$\,kpc. These values have been chosen to bracket the observed present-day core radius of Crater II, which is  $\sim 0.7$\,kpc.
These three stellar models are hereafter referred to as E1, E2, and E3 respectively, and their parameters are summarised in Table~\ref{tab:CraterII_structure}.  Core radii are related to the Einasto scale radius, $r_{\rm E}$, by the relation $R_{\rm c}\approx 1.24 \, r_{\rm E}$. In practice, stars are modelled\footnote{For details see the publicly available implementation of \citet{Errani2020}: \url{https://github.com/rerrani/nbopy}} by attaching a probability to each dark matter particle, following  the appropriate distribution function \citep[see; e.g.,][]{Bullock2005}. Using this approach, the probability of each $N$-body particle to represent a star relates the dark matter and stellar energy distributions as \begin{equation}P(E)\propto{(\mathrm{d}N/\mathrm{d}E)_\star}/{(\mathrm{d}N/\mathrm{d}E)}. 
\end{equation}
The underlying energy distributions are computed numerically as outlined in \citet{Errani2020}. 
These probabilities are computed at infall and followed throughout the orbit, where at any point the stellar structure and kinematics may be recovered from the dark matter distribution by applying the individual stellar probabilities as weights.

\subsection{Simulation code}
\label{sec:code} 
We use $10^7$-particle \textit{N}-body realisations of NFW halos, computed using the \texttt{Zeno}\footnote{\url{http://www.ifa.hawaii.edu/faculty/barnes/zeno/}} software package developed by Joshua Barnes at the University of Hawaii. This software uses Monte Carlo sampling of a given distribution function to generate systems in virial equilibrium.

We allow the halo to fully relax prior to introducing it into the MW potential, by running it first  in isolation for 5\,Gyr using the publicly available \texttt{GADGET-2} simulation code \citep{Springel2005}. Once the halo has relaxed, the $N$-body model is evolved on each of the orbits described in section \ref{sec:orbits} and in the potential of Section~\ref{sec:host} for $\sim10$\,Gyr. Forces between particles are smoothed with a Plummer-equivalent softening length of $\epsilon_\mathrm{P} = 7$\,pc and we consider our results converged outside $r_{\rm conv}=84$ pc\footnote{We define the radius of convergence, $r_{\rm conv}$, as the innermost radius where the initial circular velocities deviate by less than $\sim1\%$ from the target NFW profile.}.

%%%%%%%%%%%%%%%%%%
%
\section{Results}
\label{sec:results}

Having introduced our numerical setup, we examine next the tidal evolution of Crater II. We discuss first the evolution of the dark matter (Sec.~\ref{sec:DM}), before discussing the evolution of embedded stellar components in Sec.~\ref{sec:stars}.

\subsection{Tidal effects on  the dark matter component}
\label{sec:DM}

%%%%%%%%%%%%
\begin{figure}
\includegraphics[width=1\columnwidth]{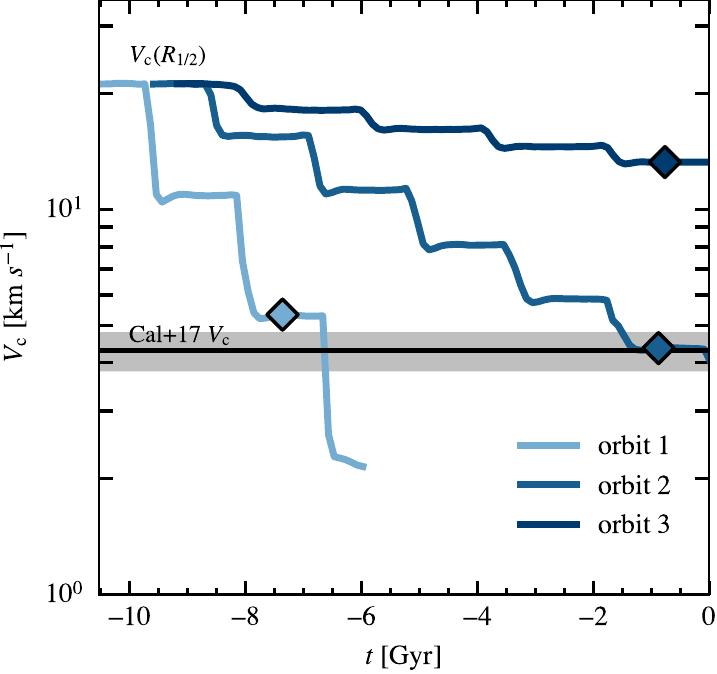}
\caption{Evolution of the circular velocity at the estimated half-light radius of Crater II (i.e., $V_\mathrm{c}$ at $r=R_{1/2}=1.066\, {\rm kpc}$) for our halo model in each of the three different orbits explored in this work; lighter shades correspond to smaller pericentres. The horizontal black solid line and grey shaded region represent the observational constraint on $V_\mathrm{c}$, as well as the $\pm 1 \sigma$ uncertainty interval, taken from \citet{Caldwell2017}. Diamonds indicate the snapshot(s) we identify for comparison with observational data. Note that, on orbit 3, our chosen halo does not get stripped enough to bring the system into agreement with the observed constraint after orbiting for $\sim 10$\,Gyr.}
\label{fig:vc}
\end{figure}
%%%%%%%%%%%%

The evolution of the circular velocity within $R_{1/2}=1.066$\,kpc of our Crater II halo models is shown in Fig.~\ref{fig:vc}. Different colors correspond to different orbits. A horizontal black line indicates, for reference, the estimated circular velocity from \citet{Caldwell2017}.

As tides strip the system, the circular velocity decreases continuously, with abrupt drops corresponding to subsequent pericentric passages. The magnitude of the decrease is heavily dependent on the assumed pericentric distance. We see that for a pericentre as large as $\sim 40$\,kpc (orbit 3), tidal effects are not enough to reduce the circular velocity enough to match the observational estimate. 

The mass loss on orbits 1 and 2, with pericentres of $4$ and $16$\,kpc, respectively, seems large enough to bring the initial halo into agreement with the observed estimate in less than $10$\,Gyr. This happens after only $2$ pericentric passages for orbit 1, but it takes $5$ full orbits for orbit 2. We note that bringing the assumed halo into agreement with the observed $V_\mathrm{c}$ implies a dramatic amount of mass loss: only $0.2\%$ of the initial dark mass remains bound when $V_\mathrm{c}(R_{1/2})$ approaches the observed value of $\sim 4$\kms.
  
We conclude that it is in principle possible to explain the unusually low velocity of Crater II if the dSph is placed on orbits with pericentric distances of order $\sim 15$\,kpc or less. This value is consistent with current observational estimates, though tighter constraints on the allowed pericentres should be able to rule out this scenario if larger pericentres are found to be favoured.

\subsection{Tidal effects on the stellar component}
\label{sec:stars}

The stellar components are also affected by tidal losses, to an extent that depends on the assumed initial density profile and  radial segregation of stars relative to the dark matter \citep[see][for a more detailed discussion]{Errani2021b}. We focus here on models E1, E2, and E3 -- exponential spheres which differ mainly in their initial core/half-light radii (see Sec.~\ref{sec:cra2_model} for details). We limit our analysis to the evolution on orbit 2, but very similar results are obtained for orbit 1, albeit on a compressed timescale.

\begin{figure}
	\includegraphics[width=1\columnwidth]{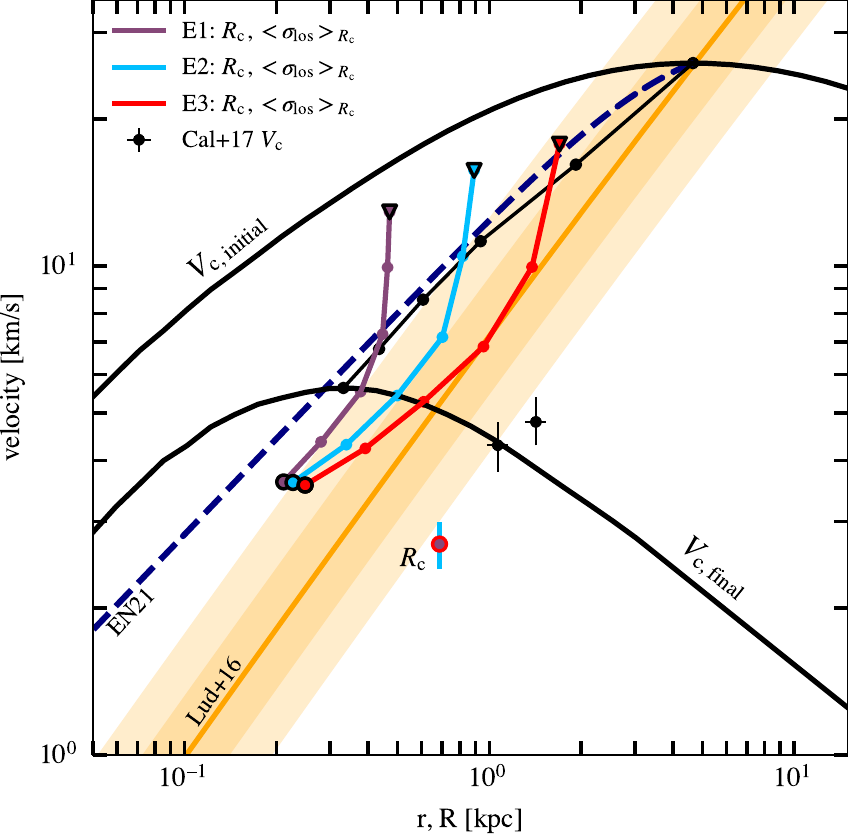}
        \caption{Evolution of the core radius, $R_{\rm c}$, and line-of-sight velocity dispersion, $\sigma_{\rm los}$, for three different stellar component models (E1, E2, E3) on orbit 2. Triangles indicate the initial values, and coloured circles the values at the final snapshot. The coloured lines trace the evolution of these parameters as a function of time; each small circle marks subsequent apocentric passages. The orange line shows the $r_{\rm max}$-$V_{\rm max}$ relation (plus $1\sigma$ and $2\sigma$ scatter bands in concentration) expected for LCDM halos at $z=0$ \citep[][in orange]{Ludlow2016}. The initial and final circular velocity of the assumed Crater II progenitor halo is shown by the solid black curves. Black circles with error bars show the observed $V_\mathrm{c}$ constraint from \citet{Caldwell2017}. The blue-red circle indicates the observed values of $R_{\rm c}$ and $\sigma_{\rm los}$ for Crater II. The dashed curve shows the ``tidal track'' of \citet{EN21}, which traces quite well the evolution of $r_{\rm max}$ and $V_{\rm max}$ of the halo (shown as a solid black line with circles at each apocentric passage) as a function of time.}
    \label{fig:tracks}
  \end{figure}

We begin by tracking the evolution of the core radius, $R_{\rm c}$, and of the line-of-sight velocity dispersion, $\sigma_{\rm los}$, of the stars (averaged within $R_{\rm c}$) in Fig.~\ref{fig:tracks}. Results for E1, E2, and E3 are shown with different coloured tracks, starting with the initial conditions (triangles) and ending, after $5$ pericentric passages, on the coloured circles highlighted in black. Each small circle along the tracks indicates a subsequent apocentric passage. The ``target'' Crater II core radius and velocity dispersion inside the core radius are shown by a red circle with a blue error bar.

Interestingly, none of the stellar tracks seems to approach the observed location of Crater II in the $R_{\rm c}$-$\sigma_{\rm los}$ plane. As the system loses mass, the velocity dispersions decrease, but so do the core radii of the stars. By the time the velocities approach the observed value, the core radii are almost $4$ times smaller than observed. This result applies to all $3$ stellar models, regardless of their initial radius.

\begin{figure*}
	\includegraphics[width=1\textwidth]{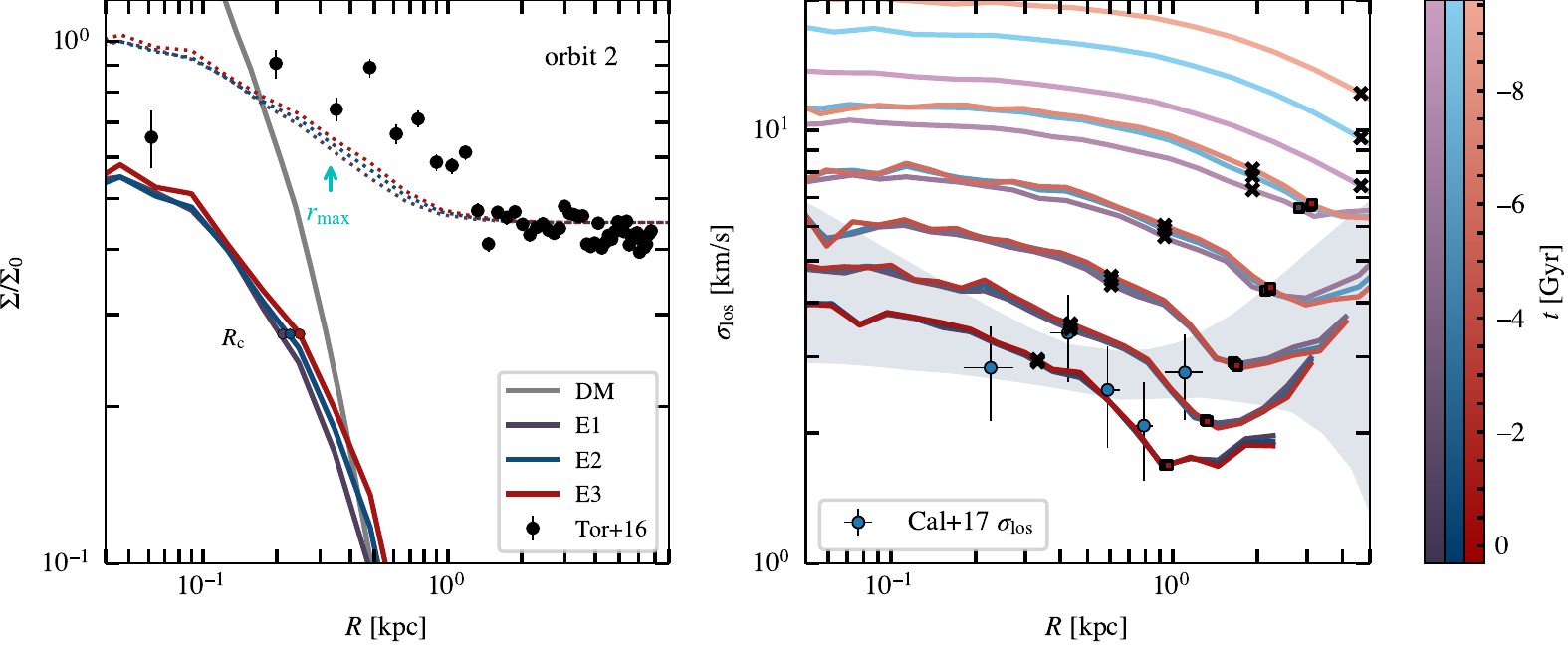}
        \caption{{\it Left:} Surface density profile of the bound remnant of stellar component models E1, E2, and E3, at the final snapshot identified for orbit 2, normalized to match the observed central density, $\Sigma_0$. Note that the core radii are much smaller than observed, despite the wide range of initial core radii ($\sim 400$ pc to $\sim 2$\,kpc) sampled by the E1, E2, and E3 models. The characteristic radius of the remnant halo, $r_{\rm max}$, shown by an arrow, is a good indicator of the tidal truncation radius of the system. The dark matter halo projected density is drawn for reference in grey.  {\it Right:} Line-of-sight velocity dispersion profiles for the bound constituents of the E1, E2, E3 models, at each subsequent apocentric passage (see colour coding, where $t=0$ is the ``final'' snapshot of orbit 2). Model velocities roughly match the observed values at the final time (bottom profiles), although the sharp tidal truncation results in a sharp decline in $\sigma_{\rm los}$ outside $r_{\rm max}$ (marked with an ``X'' on each curve). The upturn in the outer regions is due to escaping stars moving away from the remnant. Squared symbols in each profile identify the location of that feature with the radius where the local crossing time equals the time elapsed since pericentre. See text for more details. Blue circles are observed values of $\sigma_{\rm los}$ for Crater II, taken from \citet{Caldwell2017}.}
    \label{fig:profiles}
  \end{figure*}

Because of this discrepancy, reproducing the observed surface density profile of stars in Crater II is not possible, as shown in the left-hand panel of Fig.~\ref{fig:profiles}. The solid curves in that figure indicate the final density profile of bound stars in orbit 2, normalized to match approximately the central observed value for Crater II with background. Dotted curves in Fig.~\ref{fig:profiles} show the same profiles after adding a constant background of $\Sigma_{\rm bck}=0.45 \Sigma_0$. The bound stellar remnant  has a similar size in all cases, and is clearly too small to match the observed profile for Crater II.

This result is general, and not just particular to our choices of density profile, or initial radial segregation. This may be seen by noticing that, after substantial mass loss, the stellar tracks of all three systems follow the tidal track of the remnant dark matter component, indicated by the dashed black line in Fig.~\ref{fig:tracks}.This track follows the evolution of the remnant's $r_{\rm max}$ and $V_{\rm max}$ characteristic parameters which,  by the end of the simulation on orbit 2, has seen its $V_{\rm max}$ decrease by a factor of $5$ and its $r_{\rm max}$ by a factor of $14$. More importantly, as discussed by \citet{EN21}, the remnant subhalo approaches a mass profile that is well approximated by an ``exponentially truncated cusp'', with $r_{\rm max}$ roughly delineating a ``tidal truncation'' radius. 

The tidal truncation is easy to spot in the right-hand panel of Fig.~\ref{fig:profiles}, where we plot the line-of-sight velocity dispersion profiles of the bound stellar remnants at various times during the evolution. Velocity dispersions are roughly constant in the inner regions, but drop sharply outside $r_{\rm max}$, the location of which is marked with an ``X'' symbol on each curve. The decline is reversed and the velocity dispersion rises in the outermost parts because of the presence of weakly-bound stars moving radially outward. The position of this upturn approximately coincides with the location where the local crossing time equals the time elapsed since the previous pericentric passage, as discussed by \citet{Penarrubia2009}. The locations of this ``break radius'', computed\footnote{Note that the value of the constant $C=0.35$ used here differs from the $C=0.55$ value of Eq.5 in \citet{Penarrubia2009} because in that paper $R_\mathrm{b}$ refers to the location of an upturn in the surface brightness profile, whereas here we refer to an upturn in the $\sigma_{\rm los}$ profile.} as $R_\mathrm{b}=C\, \sigma_0 \, (t-t_\mathrm{p})$ (where $C=0.35$ and $\sigma_0$ is the central line of sight velocity dispersion computed as the average within $1/2$ $r_\mathrm{max}$), are indicated by small squares on each of the curves shown in the right-hand panel of Fig.~\ref{fig:profiles}.

In other words, in these late stages of tidal stripping, the size of the stellar component cannot exceed $r_{\rm max}$ because there is little bound mass left outside $r_{\rm max}$ \citep[see, also,][]{Kravtsov2010,EN21}. Indeed, an exponential cusp profile has only $\sim 13\%$ of mass outside $2\, r_{\rm max}$, and just $\sim 3\%$ outside $3\, r_{\rm max}$. This also readily explains why all three stellar models converge to a remnant with the same $R_{\rm c}$ and $\sigma_{\rm los}$, regardless of their initial size.  We emphasize that this is a general result; the results for other orbits are shown in Fig.~\ref{fig:tracks-orbits123} and results for an alternate density profile in Fig.~\ref{fig:tracks-alpha0p5}. We discuss next our results in the context of earlier work, as well as the implications of our findings for the interpretation of satellites like Crater II in a cosmological context.

\subsection{Comparison with earlier work}
\label{sec:comp_work}

The findings discussed above seem to contradict the results of \citet{Frings2017} and of  \citet{Applebaum2021}, who used simulations of tidally stripped satellites in a cosmological context to argue that it is possible to form Crater II-like satellites in LCDM cosmological simulations. We examine the reasons for the disagreement next, but emphasize that the discussion of their work relates {\it only} to their claims regarding Crater II-like systems. This discussion is not meant as wholesale criticism of their work, but rather as seeking a plausible explanation of why our conclusions differ from theirs on this particular topic.

We begin our discussion with an analysis of the results reported by  \citet{Frings2017}. Their ``satellites I and II'', in particular, seem to increase substantially in size as their halos are tidally stripped, at odds with our results. This ``tidal expansion'', however, occurs only when the orbit chosen leads to extreme tidal stripping \citep[orbit V in the notation of][]{Frings2017}.

Satellite I inhabits a halo with a pronounced ``core'' in the density profile, and thus differs from the NFW halo models studied here \citep{Maccio2017}.  For ``cored'' halos, tidal expansion of its stellar component under extreme tidal stripping is actually expected \citep[see; e.g.,][]{EPT15,Sanders2018}. Furthermore, satellite I is a fairly massive satellite with a final velocity dispersion of $\sim 8$\kms, much higher than Crater II, so there is no obvious conflict with our conclusions.

Satellite II, on the other hand, is less massive, and its halo has a less obvious core after being subjected to tides. This satellite also ``tidally expands'' when subject to extreme stripping (i.e., orbit V), leaving behind a stellar remnant with $M_\star\sim 2.1\times 10^4$\,M$_\odot$, $\sigma_{\rm los}\sim 2.5$\kms, and a large half-light radius, $r_{\rm h}\sim 1.2$\,kpc \citep[see Fig.~4, 9 in][]{Frings2017}. At face value, these properties are comparable to the properties of Crater II, and the large size\footnote{We note that ``matching the half-light radius'' is no guarantee that the actual density profile will be matched. Including weakly bound particles would result in large estimates of $r_{\rm h}$ but it would still fail to fit the density profile, unless the core radius is matched too. See discussion in Sec.~\ref{sec:observations}.}, in particular, seems at odds with our results.

However, satellite II on orbit V is stripped so severely that the reported structural properties of the stellar component are unlikely to be robust. Evidence for this comes from apparent inconsistencies between the properties of the stellar and dark matter components. For example, the total circular velocity at the stellar half-light radius is only $\sim 1$\kms \citep[see Fig.~11 in][]{Frings2017}, over a factor of $4$ smaller than the $V_{\rm c}(r_{\rm h})\sim \sqrt{3}\, \sigma_{\rm los}\approx 4.3$\kms expected from simple mass estimators \citep[see; e.g.,][]{Wolf2010}.
A discrepancy this large is not expected with or without a core, and suggests that the reported properties of the stellar component are likely affected by either limited numerical resolution or substantial departures from equilibrium. Limited numerical resolution is certainly a possibility: a circular velocity of $1$\kms implies a total mass of $\sim 2.8 \times 10^5$\,M$_\odot$ within $r_{\rm h}$, which corresponds to fewer than $\sim 200$ dark matter particles.

Another possibility is that the estimated values of $\sigma_{\rm los}$ or $r_{\rm h}$ have been inflated  by the inclusion of escaping or un-relaxed, weakly-bound particles. If this were the case, then satellite II could not be compared with Crater II, whose inner regions are likely in dynamical equilibrium (see our earlier discussion). Unfortunately, there are not enough details in \citet{Frings2017} to fully track down the origin of the discrepancy with our results, but we stand by our conclusion that Crater II's properties are inconsistent with those of a stellar component inhabiting the equilibrium tidal remnant of a cuspy NFW halo.

Limited resolution also helps to explain the findings of \citet{Applebaum2021}, who report a number of simulated satellites with stellar mass ($\sim 10^5$\,M$_\odot$), velocity dispersion ($\sim 3$\kms), and size ($\sim 1$\,kpc) comparable to Crater II, again apparently at odds with our findings. Cores cannot be invoked as an explanation here, since, according to these authors, cored halo profiles are only produced in their simulations in systems with $M_\star > 10^7$\,M$_\odot$.

\begin{figure}
\includegraphics[width=1\columnwidth]{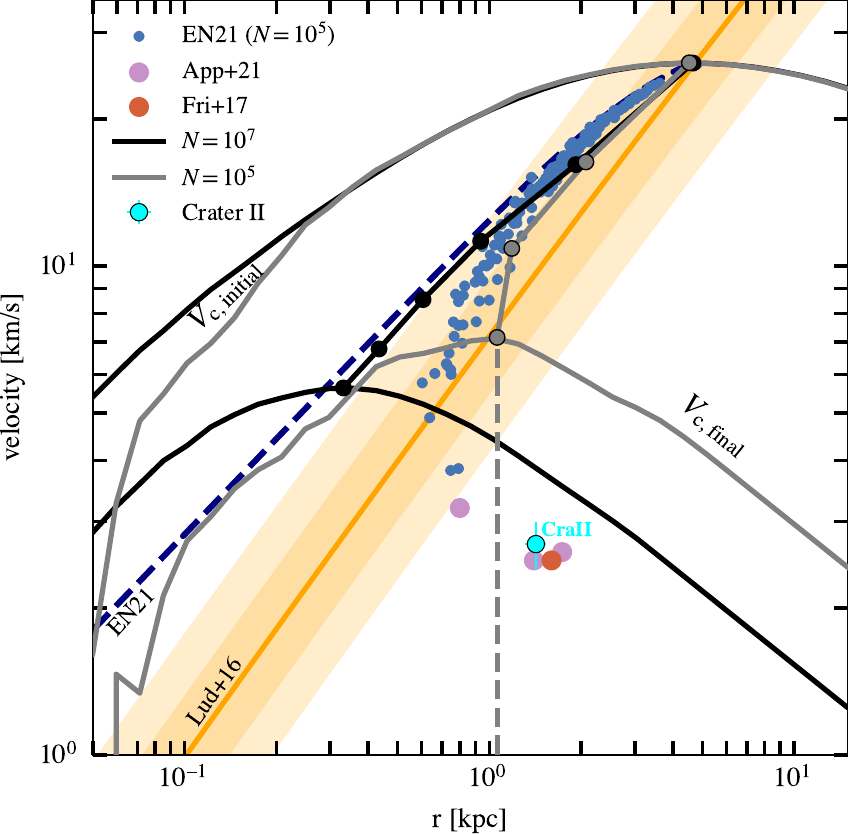}
\caption{A comparison of tidal tracks of $10^5$- and $10^7$-particle realizations of the same NFW halo (solid grey and black curves, respectively). Circles along each track mark subsequent apocentric passages on orbit 2. Also shown are the quoted values of $r_{1/2}$ and $\sigma_{\rm los}$ from \citet{Applebaum2021} (purple) and \citet{Frings2017} (orange) for Crater II-like candidates, as well as those for Crater II (\citet{Caldwell2017}; cyan symbol with error bar).  Note that unlike the $10^7$-particle halo, the $10^5$-particle halo deviates from the EN21 tidal track and disrupts fully before its fourth apocentric passage.  This deviation implies that lower resolution favours spuriously large values of $r_\mathrm{max}$ of order $\sim 1$ kpc, similar to those of the simulated Crater II-like candidates. This is confirmed by the  results of \citet{EN21} for $10^5$-particle halos (see their Fig.~A1), shown by the blue circles. See text for further details.}
\label{fig:tracks-lowres}
\end{figure}

A more likely explanation is that the large sizes reported for Crater II-like satellites by these authors result because the progenitor halos of Crater II-like systems are resolved with only $\sim 10^5$ dark matter particles, nearly $100$ times fewer than in our runs.  Indeed, limited resolution yields ``tidal tracks'' that deviate systematically from the ``EN21'' track shown in Fig.~\ref{fig:tracks}, resulting in artificially large values of $r_{\rm max}$ in the case of heavy mass loss.

We show this in  Fig.~\ref{fig:tracks-lowres}, where we plot in grey the tidal track of a $10^5$-particle NFW halo otherwise identical to that used in our runs, placed on orbit 2. Solid black curves, as in Fig.~\ref{fig:tracks}, indicate the evolution of the $10^7$-particle halo. Circles along each track indicate the values of $r_{\rm max}$ and $V_{\rm max}$ at subsequent orbital apocentres. The $10^5$-particle halo follows a track that deviates from the EN21 track after the second pericentric passage, approaching a nearly constant value of $r_{\rm max}\approx 1$ kpc before (spuriously) fully disrupting before the fourth apocentre. Such deviation is well understood as a result of insufficient resolution. Indeed, the blue circles in Fig.~\ref{fig:tracks-lowres} show the result of the evolution of many $10^5$-particle NFW halos undergoing tidal stripping, taken from Fig.~A1 of \citet{EN21}. These systems systematically deviate from the ``EN21'' track to yield overestimated values of $r_{\rm max}$.

If, as discussed earlier, the size of ``tidally limited'' galaxies is dictated by $r_{\rm max}$, then this would imply a typical size of about $\sim 1$\,kpc for the remnants of $\sim 2\times 10^9$\,M$_\odot$ halos such as those studied by \citet{Applebaum2021}. The coincidence between this radius and the reported sizes of their Crater II-like candidates supports our view that the large reported sizes of those systems are unduly affected by limited numerical resolution.

Do our results then imply that Crater II's halo had a core? One problem with this interpretation is that core formation in LCDM halos usually results from vigorous inflows and outflows of gas during galaxy formation \citep{Navarro1996b,Pontzen2012,Penarrubia2012,Benitez-Llambay2019}, which are expected to occur only in galaxies more luminous than Crater II \citep{DiCintio2014}. The model of \citet{Penarrubia2012} suggests that, for the initial stellar and virial mass of Crater II adopted in the present study study, supernova feedback can produce only a very small core in the dark matter distribution of order $\sim 0.05\,r_\mathrm{max}$. 

Indeed, according to \citet{Tollet2016} sizeable cores are only expected in galaxies whose stellar mass-to-virial mass ratio exceeds a minimum value of $\sim 3\times 10^{-4}$. This corresponds to a minimum stellar mass of $\sim 10^6$\,M$_\odot$ for the progenitor halo considered here, roughly an order of magnitude higher than the stellar mass estimated for Crater II. This interpretation thus requires that the stellar component of Crater II was once much more massive than today, which seems unlikely given its low metallicity \citep[$[{\rm Fe / H]} = -2.10\pm0.08$, see; e.g., Fig.8 in][]{Ji2021}.

Finally, \citet{Fattahi2018} used the \citet{EPT15} tidal tracks to infer the properties of potential progenitors of Crater II, and concluded that Fornax-like systems stripped of more than $99\%$ of its stars could explain the unusual properties of Crater II. Those tracks, however, are applicable only for highly segregated stellar tracers and are not valid for systems as large and diffuse as Crater II. We present a detailed and updated discussion of such tracks in a separate contribution \citep{Errani2021b}, which concludes, as we do here, that  the large size of the Crater II dSph is inconsistent with that expected for an equilibrium tidal remnant of such a massive NFW halo.

\subsection{Comparison with other dwarfs}
\begin{figure}
\includegraphics[width=1\columnwidth]{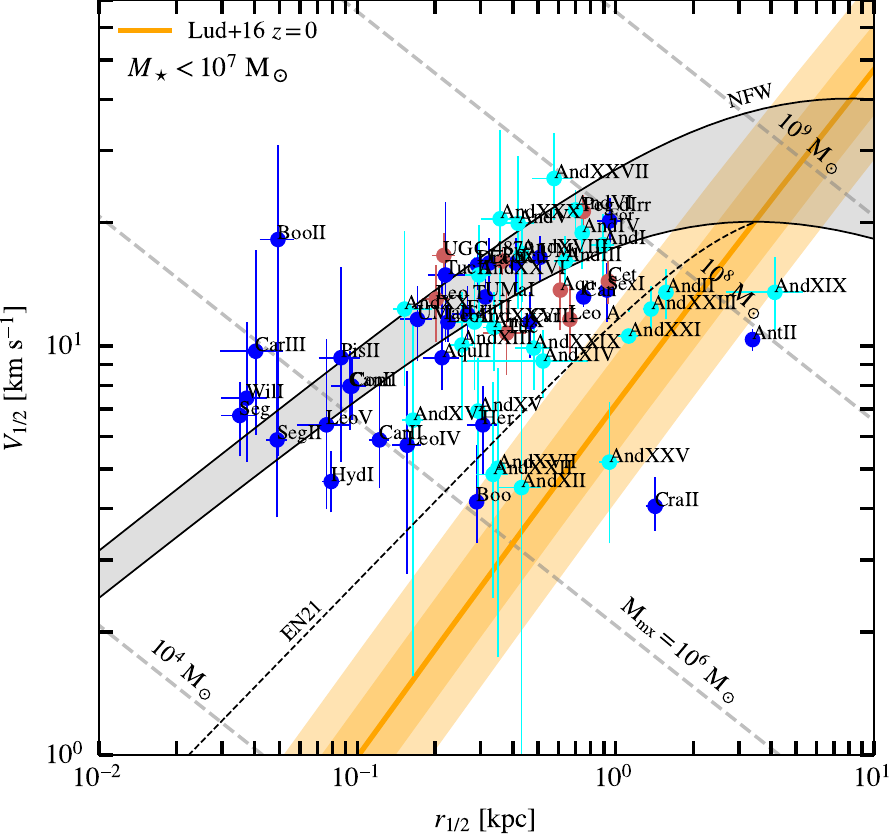}
\caption{Satellites of the Milky Way (dark blue) and of Andromeda (cyan), as well as Local Group field dwarfs (red) with $M_\star <10^7\, M_\odot$ (Fornax has also been added to the plot, for reference, although its stellar mass is slightly over the limit). 3D half-light radii, $r_{1/2}$, and circular velocities at that radius, $V_{1/2}$, are obtained using  the mass estimator of \citet{Wolf2010} applied to the projected half-light radii and line-of-sight velocity dispersions of \citet{McConnachie2012}(version January 2021), updated with recent velocity dispersions for Antlia II, Crater II \citep{Ji2021}, Tucana \citep{Taibi2020}, And XIX \citep{Collins2020}, and And XXI \citep{Collins2021}. Systems with $M_\star < 10^7\, M_\odot$ are expected to populate a narrow range of halo masses \citep{Fattahi2018}, shown schematically here by the grey band bracketing to NFW halos with $V_{\rm max}=20$ and $40$\kms and average concentrations, taken from \citet[][see orange line with $1\sigma$ and $2\sigma$ scatter bands]{Ludlow2016}. Field dwarfs and satellites unaffected by tides should fall within the grey band. Tidally affected systems can drift below the grey band, but, as discussed in our work here, should stay to the left of the tidal track (labelled ``EN21'') corresponding to that of the $V_{\rm max}=20$\kms halo, the minimum mass of a halo likely to harbour a luminous dwarf \citep[see; e.g.,][and references therein]{Benitez-Llambay2020}.
}
  \label{fig:satellites}
\end{figure}   

The structural properties of Crater II are unusual, but not unique. We compare them to other dwarf galaxies in the Local Group in Fig.~\ref{fig:satellites}, where we plot the deprojected (3D) half-light radii, $r_{1/2}$, and circular velocities at that radius, $V_{1/2}$, for Milky Way satellites (blue symbols), Andromeda satellites (cyan), as well as field dwarfs (brown).  These were estimated from the observed projected half-light radius and line-of-sight velocity dispersion using the \citet{Wolf2010} mass estimator: $r_{1/2} = (4/3) R_{1/2}$ and $V_{1/2} \approx \sqrt{3} \sigma_{\rm los}$. Data for Local Group galaxies comes from the compilation of \citet{McConnachie2012}(version January 2021), with the addition of more recent data for some dwarfs, namely Tucana, Antlia 2, Crater II, And XIX, and And XXI \citep{Collins2020,Taibi2020,Collins2021,Ji2021}.

Dark matter-dominated dwarfs inhabiting unstripped, cuspy LCDM halos with $V_{\rm max}$ in the range $(20,40)$\kms (as expected for all dwarfs with $M_\star <10^7$\,M$_\odot$, regardless how faint) should lie roughly in the area highlighted in grey  in Fig.~\ref{fig:satellites} \citep{Fattahi2018}. Remarkably, the majority of Local Group dwarfs ($\sim 70\%$) have error bars overlapping that region, which suggests that many of these galaxies indeed follow the simple expectations from LCDM simulations, with little evidence for tidal stripping playing a major role in their structure.

Systems above the grey area would suggest halos more massive than $V_{\rm max}=40$\kms, but there are no dwarfs there, except for Bootes II, which has a remarkably  high reported velocity dispersion for its size and luminosity \citep{Koch2009}. This has likely resulted from the inclusion of binary stars, which artificially inflated the estimate. Indeed, the analysis of \citet{Ji2016} concludes that Bootes II does not have a well constrained velocity dispersion, and that earlier estimates are best regarded as upper limits.

Systems well below the grey area could in principle reflect halos less massive than  $V_{\rm max}=20$\kms (marked by the bottom boundary of the grey region), but this interpretation is disfavoured because such systems would be below the minimum halo mass required for hydrogen to cool efficiently \citep{Efstathiou1992,Gnedin2000,Okamoto2009,Benitez-Llambay2020}. The most likely interpretation of such systems (all of which are MW or M31 satellites) is that their dark matter content has been reduced by tides. Our discussion above, however, indicates that tidal effects cannot push systems in NFW-like halos much below the dashed line labelled ``EN21'', which traces the ``tidal track'' of a $V_{\rm max}=20$\kms halo of average concentration.

The boundary delineated by this tidal track is a bit fuzzy because of the scatter in concentration around the mean value expected for LCDM (shown by the orange ``error bands'' in Fig.~\ref{fig:satellites}), but it is clear that there exist a number of systems whose properties seem at odds with being tidal remnants of NFW-like halos more massive than $V_{\rm max}\sim 20$\kms. Besides Crater II and Antlia II, for example, the M31 satellites And XXV and And XIX also have unusual sizes and velocity dispersions that are inconsistent with being tidal remnants of massive, cuspy LCDM halos.

The interpretation of these outliers is not completely clear, but, if they are truly dark matter-dominated equilibrium systems, reconciling their properties with LCDM requires unpalatable choices. Going through all possible options in detail is beyond the scope of this contribution, but we note that the simplest option would be to just assume that these dSphs, despite their low stellar mass, had their dark matter cusp softened or removed by baryon effects. This could allow them to reach large sizes and low velocity dispersions, thus evading the boundary set by the ``EN21'' tidal track shown in Fig.~\ref{fig:satellites}.

This solution, however,  would raise the question of why many other dwarfs show no obvious sign of such core and lie comfortably within the grey region expected for dwarfs in LCDM. What singles out some of these dwarfs, in particular, to develop such cores?  There are other alternatives, such as appealing to dark matter self-interactions to explain the presence of a ``core'' rather than a ``cusp'', but they all face a similar question: why do only a few dwarfs (and not all, or even most) seem to require a ``core'' to explain their low dark matter densities?

In this sense, the wide spread in dark matter content between dSphs shown in Fig.~\ref{fig:satellites} evokes the puzzle presented by the ``diversity'' in rotation curve shapes of dwarf irregular galaxies (dIrrs), rotationally dominated galaxies a couple of decades more luminous than the dSphs we consider here \citep{Oman2015}.

This diversity has not yet been fully explained, although it has elicited a number of proposed solutions, from baryon-induced modifications to the dark matter density profile \citep{Navarro1996b,Pontzen2012,DiCintio2014}, to modifications to the nature of cold dark matter, such as the inclusion of self-interactions \citep{Rocha2013,Ren2019}, to alterations of the laws of gravity, such as MOND \citep{McGaugh2016,Lelli2017}, to, finally, the possibility that the data has been over-interpreted, and that the diversity is driven by underestimated uncertainties in the derivation of dIrr rotation curves \citep{Oman2019}. None of these proposed alternatives seems clearly favoured at present, as discussed by \citet{Santos-Santos2020}.

\section{Summary and conclusions}
\label{SecConc}

We have used \textit{N}-body simulations to study the effect of Galactic tides on the evolution of the Crater II dSph. The main aim of this work is to probe whether tidal effects may reconcile the unusual size and kinematics of Crater II with that expected from LCDM cosmological hydrodynamical simulations. Our model assumes (i) that Crater II formed as a dark matter-dominated  stellar system initially embedded in a cuspy NFW halo; (ii) that  Crater II's halo has initially a maximum circular velocity of $\sim 26$\kms and average concentration, as expected from recent LCDM cosmological hydrodynamical simulations of the Local Group; and (iii) that Crater II stars may be modelled as spherical system with no net rotation. Our main conclusions may be summarized as follows:

\begin{itemize}

\item[(i)] The half-light radius of Crater II is poorly constrained due
  to the uncertain shape of the outer density profile. We therefore
  use the core radius when comparing simulation results against
  observations, which is less affected by assumptions about the outer
  profile shape.
   
\item[(ii)] The pericentric distance of Crater II on its orbit around
  the Milky Way is not well constrained. We use the latest available data
  on the sky position, distance, radial velocity, and proper motions of
  Crater II, together with a Galactic potential model that matches the
  observed circular velocity at the solar radius, to find that the
  allowed pericentric distance lies roughly between $\sim 10$ and
  $\sim 50$\,kpc.

\item[(iii)] The negative Galactocentric radial velocity of Crater II, together with its large distance, imply that Crater II is just past its latest apocentric passage. More importantly, this also implies that the latest pericentric passage occurred roughly $\sim 1$ Gyr ago. This implies that the inner $\sim 1$ kpc of Crater II should be close to dynamical equilibrium, excluding models where its unusual properties are ascribed to substantial departures from equilibrium.
  
\item[(iv)] Assuming that Crater II initially formed in an NFW halo
  with virial mass $M_{200}=2.7\times10^9$\,M$_\odot$ (or
  $V_{\rm max}\approx 26 $\kms, as suggested by results of the
  APOSTLE suite of LCDM Local Group simulations), we find that
  Galactic tides are able to reduce, in a Hubble time,  the characteristic velocity of the
  remnant to $\sim 6$\kms (as required to match Crater II's low
  $\sigma_{\rm los}$) if placed on an orbit with
  $r_{\rm peri}\lesssim 15$\,kpc.
   
\item[(v)] The stellar components of the halo remnants
  tidally stripped as described above have core radii of order
  $\sim 200$\,pc -- almost
  a factor of $4$ times smaller than observed. This is a general
  result for stellar components of NFW halo remnants, regardless of
  their initial radial extent. Crater II's unusual size and velocity
  dispersion are thus inconsistent with the tidal evolution of a dwarf
  galaxy in cuspy NFW halos.
  
 \item[(vi)]  Crater II's unusual properties are shared by other
   satellite galaxies in the Local Group, like Antlia II, And XXV and
   And XXI. None of these galaxies are consistent with a tidal
   interpretation in the standard LCDM scenario.

\end{itemize}

Reconciling Crater II-like systems with LCDM requires therefore some additional assumptions, none of them particularly appealing. The least unpalatable is that Crater II's halo had a shallower cusp or inner density core formed during the assembly of the galaxy. It remains to be seen whether this is possible given the low stellar mass of the dwarf. Another possibility is that Crater II's unusual properties arise because the system is quite far from equilibrium. This seems unlikely, given that Crater II is past its latest apocentric passage, and therefore its main body should have relaxed to equilibrium.

A further option is that the estimated photometric parameters of Crater II are somehow in error. This is also unlikely, given that the parameters have been estimated by independent groups \citep[see; e.g., the recent work of][]{Moskowitz2020}. Still, both these authors and the original \citet{Torrealba2016} discovery paper report a projected density profile that has an unexpected central {\it minimum} and a dynamic range of just a factor of $\sim 2$ in density. These limitations urge further work to rule out the possibility that Crater II's core radius could be perhaps substantially smaller than reported so far.

Should none of these possibilities pan out, one would be forced to consider other alternatives, such as a halo of much lower mass and concentration than assumed in our work, or even the possibility that Crater II's unusual properties signal the need to consider modifying one or more of the foundational assumptions of the LCDM paradigm.

\section*{Acknowledgements}
RE acknowledges support provided by a CITA National Fellowship and by funding from the European Research Council (ERC) under the European Unions Horizon 2020 research and innovation programme (grant agreement No. 834148). 
AF is supported by a UKRI Future Leaders Fellowship (grant no MR/T042362/1).
This work used the DiRAC@Durham facility managed by the Institute for Computational Cosmology on behalf of the STFC DiRAC HPC Facility (\url{www.dirac.ac.uk}). The equipment was funded by BEIS capital funding via STFC capital grants ST/K00042X/1, ST/P002293/1, ST/R002371/1 and ST/S002502/1, Durham University and STFC operations grant ST/R000832/1. DiRAC is part of the National e-Infrastructure.

%%%%%%%%%%%%%%%%%%%%%%%%%%%%%%%%%%%%%%%%%%%%%%%%%%
\section*{Data Availability}
The data underlying this article will be shared on reasonable request to the corresponding author.
\bibliographystyle{mnras}
\bibliography{bibliography}

\begin{thebibliography}{}
\makeatletter
\relax
\def\mn@urlcharsother{\let\do\@makeother \do\$\do\&\do\#\do\^\do\_\do\%\do\~}
\def\mn@doi{\begingroup\mn@urlcharsother \@ifnextchar [ {\mn@doi@}
  {\mn@doi@[]}}
\def\mn@doi@[#1]#2{\def\@tempa{#1}\ifx\@tempa\@empty \href
  {http://dx.doi.org/#2} {doi:#2}\else \href {http://dx.doi.org/#2} {#1}\fi
  \endgroup}
\def\mn@eprint#1#2{\mn@eprint@#1:#2::\@nil}
\def\mn@eprint@arXiv#1{\href {http://arxiv.org/abs/#1} {{\tt arXiv:#1}}}
\def\mn@eprint@dblp#1{\href {http://dblp.uni-trier.de/rec/bibtex/#1.xml}
  {dblp:#1}}
\def\mn@eprint@#1:#2:#3:#4\@nil{\def\@tempa {#1}\def\@tempb {#2}\def\@tempc
  {#3}\ifx \@tempc \@empty \let \@tempc \@tempb \let \@tempb \@tempa \fi \ifx
  \@tempb \@empty \def\@tempb {arXiv}\fi \@ifundefined
  {mn@eprint@\@tempb}{\@tempb:\@tempc}{\expandafter \expandafter \csname
  mn@eprint@\@tempb\endcsname \expandafter{\@tempc}}}

\bibitem[\protect\citeauthoryear{{Aguilar} \& {White}}{{Aguilar} \&
  {White}}{1986}]{Aguilar1986}
{Aguilar} L.~A.,  {White} S.~D.~M.,  1986, \mn@doi [\apj] {10.1086/164396},
  \href {https://ui.adsabs.harvard.edu/abs/1986ApJ...307...97A} {307, 97}

\bibitem[\protect\citeauthoryear{{Amorisco}}{{Amorisco}}{2019}]{Amorisco2019}
{Amorisco} N.~C.,  2019, \mn@doi [\mnras] {10.1093/mnrasl/slz121}, \href
  {https://ui.adsabs.harvard.edu/abs/2019MNRAS.489L..22A} {489, L22}

\bibitem[\protect\citeauthoryear{{Applebaum}, {Brooks}, {Christensen},
  {Munshi}, {Quinn}, {Shen}  \& {Tremmel}}{{Applebaum}
  et~al.}{2021}]{Applebaum2021}
{Applebaum} E.,  {Brooks} A.~M.,  {Christensen} C.~R.,  {Munshi} F.,  {Quinn}
  T.~R.,  {Shen} S.,   {Tremmel} M.,  2021, \mn@doi [\apj]
  {10.3847/1538-4357/abcafa}, \href
  {https://ui.adsabs.harvard.edu/abs/2021ApJ...906...96A} {906, 96}

\bibitem[\protect\citeauthoryear{{Benitez-Llambay} \&
  {Frenk}}{{Benitez-Llambay} \& {Frenk}}{2020}]{Benitez-Llambay2020}
{Benitez-Llambay} A.,  {Frenk} C.,  2020, \mn@doi [\mnras]
  {10.1093/mnras/staa2698}, \href
  {https://ui.adsabs.harvard.edu/abs/2020MNRAS.498.4887B} {498, 4887}

\bibitem[\protect\citeauthoryear{{Ben{\'\i}tez-Llambay}, {Frenk}, {Ludlow}  \&
  {Navarro}}{{Ben{\'\i}tez-Llambay} et~al.}{2019}]{Benitez-Llambay2019}
{Ben{\'\i}tez-Llambay} A.,  {Frenk} C.~S.,  {Ludlow} A.~D.,   {Navarro} J.~F.,
  2019, \mn@doi [\mnras] {10.1093/mnras/stz1890}, \href
  {https://ui.adsabs.harvard.edu/abs/2019MNRAS.488.2387B} {488, 2387}

\bibitem[\protect\citeauthoryear{{Borukhovetskaya}, {Errani}, {Navarro},
  {Fattahi}  \& {Santos-Santos}}{{Borukhovetskaya}
  et~al.}{2022}]{Borukhovetskaya2021}
{Borukhovetskaya} A.,  {Errani} R.,  {Navarro} J.~F.,  {Fattahi} A.,
  {Santos-Santos} I.,  2022, \mn@doi [\mnras] {10.1093/mnras/stab2912}, \href
  {https://ui.adsabs.harvard.edu/abs/2022MNRAS.509.5330B} {509, 5330}

\bibitem[\protect\citeauthoryear{{Bullock} \& {Johnston}}{{Bullock} \&
  {Johnston}}{2005}]{Bullock2005}
{Bullock} J.~S.,  {Johnston} K.~V.,  2005, \mn@doi [\apj] {10.1086/497422},
  \href {https://ui.adsabs.harvard.edu/abs/2005ApJ...635..931B} {635, 931}

\bibitem[\protect\citeauthoryear{{Caldwell} et~al.,}{{Caldwell}
  et~al.}{2017}]{Caldwell2017}
{Caldwell} N.,  et~al., 2017, \mn@doi [\apj] {10.3847/1538-4357/aa688e}, \href
  {https://ui.adsabs.harvard.edu/abs/2017ApJ...839...20C} {839, 20}

\bibitem[\protect\citeauthoryear{{Collins}, {Tollerud}, {Rich}, {Ibata},
  {Martin}, {Chapman}, {Gilbert}  \& {Preston}}{{Collins}
  et~al.}{2020}]{Collins2020}
{Collins} M. L.~M.,  {Tollerud} E.~J.,  {Rich} R.~M.,  {Ibata} R.~A.,  {Martin}
  N.~F.,  {Chapman} S.~C.,  {Gilbert} K.~M.,   {Preston} J.,  2020, \mn@doi
  [\mnras] {10.1093/mnras/stz3252}, \href
  {https://ui.adsabs.harvard.edu/abs/2020MNRAS.491.3496C} {491, 3496}

\bibitem[\protect\citeauthoryear{{Collins} et~al.,}{{Collins}
  et~al.}{2021}]{Collins2021}
{Collins} M. L.~M.,  et~al., 2021, \mn@doi [\mnras] {10.1093/mnras/stab1624},
  \href {https://ui.adsabs.harvard.edu/abs/2021MNRAS.505.5686C} {505, 5686}

\bibitem[\protect\citeauthoryear{{Di Cintio}, {Brook}, {Dutton}, {Macci{\`o}},
  {Stinson}  \& {Knebe}}{{Di Cintio} et~al.}{2014}]{DiCintio2014}
{Di Cintio} A.,  {Brook} C.~B.,  {Dutton} A.~A.,  {Macci{\`o}} A.~V.,
  {Stinson} G.~S.,   {Knebe} A.,  2014, \mn@doi [\mnras]
  {10.1093/mnras/stu729}, \href
  {http://adsabs.harvard.edu/abs/2014MNRAS.441.2986D} {441, 2986}

\bibitem[\protect\citeauthoryear{{Efstathiou}}{{Efstathiou}}{1992}]{Efstathiou1992}
{Efstathiou} G.,  1992, \mn@doi [\mnras] {10.1093/mnras/256.1.43P}, \href
  {http://adsabs.harvard.edu/abs/1992MNRAS.256P..43E} {256, 43P}

\bibitem[\protect\citeauthoryear{{Einasto}}{{Einasto}}{1965}]{Einasto1965}
{Einasto} J.,  1965, Trudy Astrofizicheskogo Instituta Alma-Ata, \href
  {https://ui.adsabs.harvard.edu/abs/1965TrAlm...5...87E} {5, 87}

\bibitem[\protect\citeauthoryear{{Errani} \& {Navarro}}{{Errani} \&
  {Navarro}}{2021}]{EN21}
{Errani} R.,  {Navarro} J.~F.,  2021, \mn@doi [\mnras]
  {10.1093/mnras/stab1215}, \href
  {https://ui.adsabs.harvard.edu/abs/2021MNRAS.505...18E} {505, 18}

\bibitem[\protect\citeauthoryear{{Errani} \& {Pe{\~n}arrubia}}{{Errani} \&
  {Pe{\~n}arrubia}}{2020}]{Errani2020}
{Errani} R.,  {Pe{\~n}arrubia} J.,  2020, \mn@doi [\mnras]
  {10.1093/mnras/stz3349}, \href
  {https://ui.adsabs.harvard.edu/abs/2020MNRAS.491.4591E} {491, 4591}

\bibitem[\protect\citeauthoryear{{Errani}, {Pe{\~n}arrubia}  \&
  {Tormen}}{{Errani} et~al.}{2015}]{EPT15}
{Errani} R.,  {Pe{\~n}arrubia} J.,   {Tormen} G.,  2015, \mn@doi [\mnras]
  {10.1093/mnrasl/slv012}, \href
  {http://adsabs.harvard.edu/abs/2015MNRAS.449L..46E} {449, L46}

\bibitem[\protect\citeauthoryear{{Errani}, {Navarro}, {Ibata}  \&
  {Pe{\~n}arrubia}}{{Errani} et~al.}{2022}]{Errani2021b}
{Errani} R.,  {Navarro} J.~F.,  {Ibata} R.,   {Pe{\~n}arrubia} J.,  2022,
  \mn@doi [\mnras] {10.1093/mnras/stac476}, \href
  {https://ui.adsabs.harvard.edu/abs/2022MNRAS.511.6001E} {511, 6001}

\bibitem[\protect\citeauthoryear{{Fattahi} et~al.,}{{Fattahi}
  et~al.}{2016}]{Fattahi2016a}
{Fattahi} A.,  et~al., 2016, \mn@doi [\mnras] {10.1093/mnras/stv2970}, \href
  {https://ui.adsabs.harvard.edu/abs/2016MNRAS.457..844F} {457, 844}

\bibitem[\protect\citeauthoryear{{Fattahi}, {Navarro}, {Frenk}, {Oman},
  {Sawala}  \& {Schaller}}{{Fattahi} et~al.}{2018}]{Fattahi2018}
{Fattahi} A.,  {Navarro} J.~F.,  {Frenk} C.~S.,  {Oman} K.~A.,  {Sawala} T.,
  {Schaller} M.,  2018, \mn@doi [\mnras] {10.1093/mnras/sty408}, \href
  {https://ui.adsabs.harvard.edu/abs/2018MNRAS.476.3816F} {476, 3816}

\bibitem[\protect\citeauthoryear{{Frings}, {Macci{\`o}}, {Buck}, {Penzo},
  {Dutton}, {Blank}  \& {Obreja}}{{Frings} et~al.}{2017}]{Frings2017}
{Frings} J.,  {Macci{\`o}} A.,  {Buck} T.,  {Penzo} C.,  {Dutton} A.,  {Blank}
  M.,   {Obreja} A.,  2017, \mn@doi [\mnras] {10.1093/mnras/stx2171}, \href
  {https://ui.adsabs.harvard.edu/abs/2017MNRAS.472.3378F} {472, 3378}

\bibitem[\protect\citeauthoryear{{Fritz}, {Battaglia}, {Pawlowski},
  {Kallivayalil}, {van der Marel}, {Sohn}, {Brook}  \& {Besla}}{{Fritz}
  et~al.}{2018}]{Fritz2018}
{Fritz} T.~K.,  {Battaglia} G.,  {Pawlowski} M.~S.,  {Kallivayalil} N.,  {van
  der Marel} R.,  {Sohn} S.~T.,  {Brook} C.,   {Besla} G.,  2018, \mn@doi
  [\aap] {10.1051/0004-6361/201833343}, \href
  {https://ui.adsabs.harvard.edu/abs/2018A&A...619A.103F} {619, A103}

\bibitem[\protect\citeauthoryear{{Fu}, {Simon}  \& {Alarc{\'o}n Jara}}{{Fu}
  et~al.}{2019}]{Fu2019}
{Fu} S.~W.,  {Simon} J.~D.,   {Alarc{\'o}n Jara} A.~G.,  2019, \mn@doi [\apj]
  {10.3847/1538-4357/ab3658}, \href
  {https://ui.adsabs.harvard.edu/abs/2019ApJ...883...11F} {883, 11}

\bibitem[\protect\citeauthoryear{{Gnedin}}{{Gnedin}}{2000}]{Gnedin2000}
{Gnedin} N.~Y.,  2000, \mn@doi [\apj] {10.1086/317042}, \href
  {http://adsabs.harvard.edu/abs/2000ApJ...542..535G} {542, 535}

\bibitem[\protect\citeauthoryear{{Hayashi}, {Navarro}, {Taylor}, {Stadel}  \&
  {Quinn}}{{Hayashi} et~al.}{2003}]{Hayashi2003}
{Hayashi} E.,  {Navarro} J.~F.,  {Taylor} J.~E.,  {Stadel} J.,   {Quinn} T.,
  2003, \mn@doi [\apj] {10.1086/345788}, \href
  {https://ui.adsabs.harvard.edu/abs/2003ApJ...584..541H} {584, 541}

\bibitem[\protect\citeauthoryear{{Hernquist}}{{Hernquist}}{1990}]{Hernquist1990}
{Hernquist} L.,  1990, \mn@doi [\apj] {10.1086/168845}, \href
  {https://ui.adsabs.harvard.edu/abs/1990ApJ...356..359H} {356, 359}

\bibitem[\protect\citeauthoryear{{Ji}, {Frebel}, {Simon}  \& {Geha}}{{Ji}
  et~al.}{2016}]{Ji2016}
{Ji} A.~P.,  {Frebel} A.,  {Simon} J.~D.,   {Geha} M.,  2016, \mn@doi [\apj]
  {10.3847/0004-637X/817/1/41}, \href
  {https://ui.adsabs.harvard.edu/abs/2016ApJ...817...41J} {817, 41}

\bibitem[\protect\citeauthoryear{{Ji} et~al.,}{{Ji} et~al.}{2021}]{Ji2021}
{Ji} A.~P.,  et~al., 2021, \mn@doi [\apj] {10.3847/1538-4357/ac1869}, \href
  {https://ui.adsabs.harvard.edu/abs/2021ApJ...921...32J} {921, 32}

\bibitem[\protect\citeauthoryear{{Kallivayalil} et~al.,}{{Kallivayalil}
  et~al.}{2018}]{Kallivayalil2018}
{Kallivayalil} N.,  et~al., 2018, \mn@doi [\apj] {10.3847/1538-4357/aadfee},
  \href {https://ui.adsabs.harvard.edu/abs/2018ApJ...867...19K} {867, 19}

\bibitem[\protect\citeauthoryear{{Koch} et~al.,}{{Koch}
  et~al.}{2009}]{Koch2009}
{Koch} A.,  et~al., 2009, \mn@doi [\apj] {10.1088/0004-637X/690/1/453}, \href
  {https://ui.adsabs.harvard.edu/abs/2009ApJ...690..453K} {690, 453}

\bibitem[\protect\citeauthoryear{{Kravtsov}}{{Kravtsov}}{2010}]{Kravtsov2010}
{Kravtsov} A.,  2010, \mn@doi [Advances in Astronomy] {10.1155/2010/281913},
  \href {https://ui.adsabs.harvard.edu/abs/2010AdAst2010E...8K} {2010, 281913}

\bibitem[\protect\citeauthoryear{{Lelli}, {McGaugh}, {Schombert}  \&
  {Pawlowski}}{{Lelli} et~al.}{2017}]{Lelli2017}
{Lelli} F.,  {McGaugh} S.~S.,  {Schombert} J.~M.,   {Pawlowski} M.~S.,  2017,
  \mn@doi [\apj] {10.3847/1538-4357/836/2/152}, \href
  {https://ui.adsabs.harvard.edu/abs/2017ApJ...836..152L} {836, 152}

\bibitem[\protect\citeauthoryear{{Ludlow}, {Bose}, {Angulo}, {Wang},
  {Hellwing}, {Navarro}, {Cole}  \& {Frenk}}{{Ludlow}
  et~al.}{2016}]{Ludlow2016}
{Ludlow} A.~D.,  {Bose} S.,  {Angulo} R.~E.,  {Wang} L.,  {Hellwing} W.~A.,
  {Navarro} J.~F.,  {Cole} S.,   {Frenk} C.~S.,  2016, \mn@doi [\mnras]
  {10.1093/mnras/stw1046}, \href
  {https://ui.adsabs.harvard.edu/abs/2016MNRAS.460.1214L} {460, 1214}

\bibitem[\protect\citeauthoryear{{Macci{\`o}}, {Frings}, {Buck}, {Penzo},
  {Dutton}, {Blank}  \& {Obreja}}{{Macci{\`o}} et~al.}{2017}]{Maccio2017}
{Macci{\`o}} A.~V.,  {Frings} J.,  {Buck} T.,  {Penzo} C.,  {Dutton} A.~A.,
  {Blank} M.,   {Obreja} A.,  2017, \mn@doi [\mnras] {10.1093/mnras/stx2048},
  \href {https://ui.adsabs.harvard.edu/abs/2017MNRAS.472.2356M} {472, 2356}

\bibitem[\protect\citeauthoryear{{McConnachie}}{{McConnachie}}{2012}]{McConnachie2012}
{McConnachie} A.~W.,  2012, \mn@doi [\aj] {10.1088/0004-6256/144/1/4}, \href
  {http://adsabs.harvard.edu/abs/2012AJ....144....4M} {144, 4}

\bibitem[\protect\citeauthoryear{{McConnachie} \& {Venn}}{{McConnachie} \&
  {Venn}}{2020}]{McConnachie2020-DR3}
{McConnachie} A.~W.,  {Venn} K.~A.,  2020, \mn@doi [Research Notes of the
  American Astronomical Society] {10.3847/2515-5172/abd18b}, \href
  {https://ui.adsabs.harvard.edu/abs/2020RNAAS...4..229M} {4, 229}

\bibitem[\protect\citeauthoryear{{McGaugh}}{{McGaugh}}{2016}]{McGaugh2016}
{McGaugh} S.~S.,  2016, \mn@doi [\apjl] {10.3847/2041-8205/832/1/L8}, \href
  {https://ui.adsabs.harvard.edu/abs/2016ApJ...832L...8M} {832, L8}

\bibitem[\protect\citeauthoryear{{McMillan}}{{McMillan}}{2011}]{McMillan2011}
{McMillan} P.~J.,  2011, \mn@doi [\mnras] {10.1111/j.1365-2966.2011.18564.x},
  \href {https://ui.adsabs.harvard.edu/abs/2011MNRAS.414.2446M} {414, 2446}

\bibitem[\protect\citeauthoryear{{Miyamoto} \& {Nagai}}{{Miyamoto} \&
  {Nagai}}{1975}]{Miyamoto1975}
{Miyamoto} M.,  {Nagai} R.,  1975, \pasj, \href
  {https://ui.adsabs.harvard.edu/abs/1975PASJ...27..533M} {27, 533}

\bibitem[\protect\citeauthoryear{{Moskowitz} \& {Walker}}{{Moskowitz} \&
  {Walker}}{2020}]{Moskowitz2020}
{Moskowitz} A.~G.,  {Walker} M.~G.,  2020, \mn@doi [\apj]
  {10.3847/1538-4357/ab7459}, \href
  {https://ui.adsabs.harvard.edu/abs/2020ApJ...892...27M} {892, 27}

\bibitem[\protect\citeauthoryear{{Navarro}}{{Navarro}}{1990}]{Navarro1990}
{Navarro} J.~F.,  1990, \mn@doi [\mnras] {10.1093/mnras/242.3.311}, \href
  {https://ui.adsabs.harvard.edu/abs/1990MNRAS.242..311N} {242, 311}

\bibitem[\protect\citeauthoryear{{Navarro}, {Eke}  \& {Frenk}}{{Navarro}
  et~al.}{1996a}]{Navarro1996b}
{Navarro} J.~F.,  {Eke} V.~R.,   {Frenk} C.~S.,  1996a, \mnras, \href
  {http://adsabs.harvard.edu/abs/1996MNRAS.283L..72N} {283, L72}

\bibitem[\protect\citeauthoryear{{Navarro}, {Frenk}  \& {White}}{{Navarro}
  et~al.}{1996b}]{Navarro1996a}
{Navarro} J.~F.,  {Frenk} C.~S.,   {White} S. D.~M.,  1996b, \mn@doi [\apj]
  {10.1086/177173}, \href
  {https://ui.adsabs.harvard.edu/abs/1996ApJ...462..563N} {462, 563}

\bibitem[\protect\citeauthoryear{{Navarro}, {Frenk}  \& {White}}{{Navarro}
  et~al.}{1997}]{Navarro1997}
{Navarro} J.~F.,  {Frenk} C.~S.,   {White} S. D.~M.,  1997, \mn@doi [\apj]
  {10.1086/304888}, \href
  {https://ui.adsabs.harvard.edu/abs/1997ApJ...490..493N} {490, 493}

\bibitem[\protect\citeauthoryear{{Okamoto} \& {Frenk}}{{Okamoto} \&
  {Frenk}}{2009}]{Okamoto2009}
{Okamoto} T.,  {Frenk} C.~S.,  2009, \mn@doi [\mnras]
  {10.1111/j.1745-3933.2009.00748.x}, \href
  {https://ui.adsabs.harvard.edu/abs/2009MNRAS.399L.174O} {399, L174}

\bibitem[\protect\citeauthoryear{{Oman} et~al.,}{{Oman}
  et~al.}{2015}]{Oman2015}
{Oman} K.~A.,  et~al., 2015, \mn@doi [\mnras] {10.1093/mnras/stv1504}, \href
  {https://ui.adsabs.harvard.edu/abs/2015MNRAS.452.3650O} {452, 3650}

\bibitem[\protect\citeauthoryear{{Oman}, {Marasco}, {Navarro}, {Frenk},
  {Schaye}  \& {Ben{\'\i}tez-Llambay}}{{Oman} et~al.}{2019}]{Oman2019}
{Oman} K.~A.,  {Marasco} A.,  {Navarro} J.~F.,  {Frenk} C.~S.,  {Schaye} J.,
  {Ben{\'\i}tez-Llambay} A.,  2019, \mn@doi [\mnras] {10.1093/mnras/sty2687},
  \href {https://ui.adsabs.harvard.edu/abs/2019MNRAS.482..821O} {482, 821}

\bibitem[\protect\citeauthoryear{{Pe{\~n}arrubia}, {Navarro}  \&
  {McConnachie}}{{Pe{\~n}arrubia} et~al.}{2008}]{penarrubia2008}
{Pe{\~n}arrubia} J.,  {Navarro} J.~F.,   {McConnachie} A.~W.,  2008, \mn@doi
  [ApJ] {10.1086/523686}, \href
  {http://adsabs.harvard.edu/abs/2008ApJ...673..226P} {673, 226}

\bibitem[\protect\citeauthoryear{{Pe{\~n}arrubia}, {Navarro}, {McConnachie}  \&
  {Martin}}{{Pe{\~n}arrubia} et~al.}{2009}]{Penarrubia2009}
{Pe{\~n}arrubia} J.,  {Navarro} J.~F.,  {McConnachie} A.~W.,   {Martin} N.~F.,
  2009, \mn@doi [\apj] {10.1088/0004-637X/698/1/222}, \href
  {https://ui.adsabs.harvard.edu/abs/2009ApJ...698..222P} {698, 222}

\bibitem[\protect\citeauthoryear{{Pe{\~n}arrubia}, {Benson}, {Walker},
  {Gilmore}, {McConnachie}  \& {Mayer}}{{Pe{\~n}arrubia}
  et~al.}{2010}]{Penarrubia2010}
{Pe{\~n}arrubia} J.,  {Benson} A.~J.,  {Walker} M.~G.,  {Gilmore} G.,
  {McConnachie} A.~W.,   {Mayer} L.,  2010, \mn@doi [\mnras]
  {10.1111/j.1365-2966.2010.16762.x}, \href
  {https://ui.adsabs.harvard.edu/abs/2010MNRAS.406.1290P} {406, 1290}

\bibitem[\protect\citeauthoryear{{Pe{\~n}arrubia}, {Pontzen}, {Walker}  \&
  {Koposov}}{{Pe{\~n}arrubia} et~al.}{2012}]{Penarrubia2012}
{Pe{\~n}arrubia} J.,  {Pontzen} A.,  {Walker} M.~G.,   {Koposov} S.~E.,  2012,
  \mn@doi [\apjl] {10.1088/2041-8205/759/2/L42}, \href
  {https://ui.adsabs.harvard.edu/abs/2012ApJ...759L..42P} {759, L42}

\bibitem[\protect\citeauthoryear{{Plummer}}{{Plummer}}{1911}]{Plummer1911}
{Plummer} H.~C.,  1911, \mn@doi [\mnras] {10.1093/mnras/71.5.460}, \href
  {https://ui.adsabs.harvard.edu/abs/1911MNRAS..71..460P} {71, 460}

\bibitem[\protect\citeauthoryear{{Pontzen} \& {Governato}}{{Pontzen} \&
  {Governato}}{2012}]{Pontzen2012}
{Pontzen} A.,  {Governato} F.,  2012, \mn@doi [\mnras]
  {10.1111/j.1365-2966.2012.20571.x}, \href
  {http://adsabs.harvard.edu/abs/2012MNRAS.421.3464P} {421, 3464}

\bibitem[\protect\citeauthoryear{{Read} \& {Gilmore}}{{Read} \&
  {Gilmore}}{2005}]{Read2005}
{Read} J.~I.,  {Gilmore} G.,  2005, \mn@doi [\mnras]
  {10.1111/j.1365-2966.2004.08424.x}, \href
  {https://ui.adsabs.harvard.edu/abs/2005MNRAS.356..107R} {356, 107}

\bibitem[\protect\citeauthoryear{{Ren}, {Kwa}, {Kaplinghat}  \& {Yu}}{{Ren}
  et~al.}{2019}]{Ren2019}
{Ren} T.,  {Kwa} A.,  {Kaplinghat} M.,   {Yu} H.-B.,  2019, \mn@doi [Physical
  Review X] {10.1103/PhysRevX.9.031020}, \href
  {https://ui.adsabs.harvard.edu/abs/2019PhRvX...9c1020R} {9, 031020}

\bibitem[\protect\citeauthoryear{{Rocha}, {Peter}, {Bullock}, {Kaplinghat},
  {Garrison-Kimmel}, {O{\~n}orbe}  \& {Moustakas}}{{Rocha}
  et~al.}{2013}]{Rocha2013}
{Rocha} M.,  {Peter} A.~H.~G.,  {Bullock} J.~S.,  {Kaplinghat} M.,
  {Garrison-Kimmel} S.,  {O{\~n}orbe} J.,   {Moustakas} L.~A.,  2013, \mn@doi
  [\mnras] {10.1093/mnras/sts514}, \href
  {http://adsabs.harvard.edu/abs/2013MNRAS.430...81R} {430, 81}

\bibitem[\protect\citeauthoryear{{Sanders}, {Evans}  \& {Dehnen}}{{Sanders}
  et~al.}{2018}]{Sanders2018}
{Sanders} J.~L.,  {Evans} N.~W.,   {Dehnen} W.,  2018, \mn@doi [\mnras]
  {10.1093/mnras/sty1278}, \href
  {http://adsabs.harvard.edu/abs/2018MNRAS.478.3879S} {478, 3879}

\bibitem[\protect\citeauthoryear{{Santos-Santos} et~al.,}{{Santos-Santos}
  et~al.}{2020}]{Santos-Santos2020}
{Santos-Santos} I. M.~E.,  et~al., 2020, \mn@doi [\mnras]
  {10.1093/mnras/staa1072}, \href
  {https://ui.adsabs.harvard.edu/abs/2020MNRAS.495...58S} {495, 58}

\bibitem[\protect\citeauthoryear{{Santos-Santos}, {Sales}, {Fattahi}  \&
  {Navarro}}{{Santos-Santos} et~al.}{2021}]{Santos-Santos2021}
{Santos-Santos} I. M.~E.,  {Sales} L.~V.,  {Fattahi} A.,   {Navarro} J.~F.,
  2021, arXiv e-prints, \href
  {https://ui.adsabs.harvard.edu/abs/2021arXiv211101158S} {p. arXiv:2111.01158}

\bibitem[\protect\citeauthoryear{{Sawala} et~al.,}{{Sawala}
  et~al.}{2016}]{Sawala2016}
{Sawala} T.,  et~al., 2016, \mn@doi [\mnras] {10.1093/mnras/stw145}, \href
  {https://ui.adsabs.harvard.edu/abs/2016MNRAS.457.1931S} {457, 1931}

\bibitem[\protect\citeauthoryear{{Springel}}{{Springel}}{2005}]{Springel2005}
{Springel} V.,  2005, \mn@doi [\mnras] {10.1111/j.1365-2966.2005.09655.x},
  \href {https://ui.adsabs.harvard.edu/abs/2005MNRAS.364.1105S} {364, 1105}

\bibitem[\protect\citeauthoryear{{Taibi}, {Battaglia}, {Rejkuba}, {Leaman},
  {Kacharov}, {Iorio}, {Jablonka}  \& {Zoccali}}{{Taibi}
  et~al.}{2020}]{Taibi2020}
{Taibi} S.,  {Battaglia} G.,  {Rejkuba} M.,  {Leaman} R.,  {Kacharov} N.,
  {Iorio} G.,  {Jablonka} P.,   {Zoccali} M.,  2020, \mn@doi [\aap]
  {10.1051/0004-6361/201937240}, \href
  {https://ui.adsabs.harvard.edu/abs/2020A&A...635A.152T} {635, A152}

\bibitem[\protect\citeauthoryear{{Tollet} et~al.,}{{Tollet}
  et~al.}{2016}]{Tollet2016}
{Tollet} E.,  et~al., 2016, \mn@doi [\mnras] {10.1093/mnras/stv2856}, \href
  {https://ui.adsabs.harvard.edu/abs/2016MNRAS.456.3542T} {456, 3542}

\bibitem[\protect\citeauthoryear{{Torrealba}, {Koposov}, {Belokurov}  \&
  {Irwin}}{{Torrealba} et~al.}{2016}]{Torrealba2016}
{Torrealba} G.,  {Koposov} S.~E.,  {Belokurov} V.,   {Irwin} M.,  2016, \mn@doi
  [\mnras] {10.1093/mnras/stw733}, \href
  {https://ui.adsabs.harvard.edu/abs/2016MNRAS.459.2370T} {459, 2370}

\bibitem[\protect\citeauthoryear{{Torrealba} et~al.,}{{Torrealba}
  et~al.}{2019}]{Torrealba2019}
{Torrealba} G.,  et~al., 2019, \mn@doi [\mnras] {10.1093/mnras/stz1624}, \href
  {https://ui.adsabs.harvard.edu/abs/2019MNRAS.488.2743T} {488, 2743}

\bibitem[\protect\citeauthoryear{\VAN{Bosch}{Van}{van den}~Bosch \&
  {Ogiya}}{\VAN{Bosch}{Van}{van den}~Bosch \& {Ogiya}}{2018}]{vdBOgiya2018}
\VAN{Bosch}{Van}{van den}~Bosch F.~C.,  {Ogiya} G.,  2018, \mn@doi [\mnras]
  {10.1093/mnras/sty084}, \href
  {http://adsabs.harvard.edu/abs/2018MNRAS.475.4066V} {475, 4066}

\bibitem[\protect\citeauthoryear{\VAN{Bosch}{Van}{van den}~Bosch, {Ogiya},
  {Hahn}  \& {Burkert}}{\VAN{Bosch}{Van}{van den}~Bosch et~al.}{2018}]{vdb2018}
\VAN{Bosch}{Van}{van den}~Bosch F.~C.,  {Ogiya} G.,  {Hahn} O.,   {Burkert} A.,
   2018, \mn@doi [\mnras] {10.1093/mnras/stx2956}, \href
  {http://adsabs.harvard.edu/abs/2018MNRAS.474.3043V} {474, 3043}

\bibitem[\protect\citeauthoryear{\VAN{Dokkum}{Van}{van}~Dokkum, {Abraham},
  {Merritt}, {Zhang}, {Geha}  \& {Conroy}}{\VAN{Dokkum}{Van}{van}~Dokkum
  et~al.}{2015}]{vanDokkum2015}
\VAN{Dokkum}{Van}{van}~Dokkum P.~G.,  {Abraham} R.,  {Merritt} A.,  {Zhang} J.,
   {Geha} M.,   {Conroy} C.,  2015, \mn@doi [\apjl]
  {10.1088/2041-8205/798/2/L45}, \href
  {https://ui.adsabs.harvard.edu/abs/2015ApJ...798L..45V} {798, L45}

\bibitem[\protect\citeauthoryear{{Walker}, {Mateo}, {Olszewski},
  {Pe{\~n}arrubia}, {Evans}  \& {Gilmore}}{{Walker} et~al.}{2009}]{Walker2009}
{Walker} M.~G.,  {Mateo} M.,  {Olszewski} E.~W.,  {Pe{\~n}arrubia} J.,  {Evans}
  N.~W.,   {Gilmore} G.,  2009, \mn@doi [\apj] {10.1088/0004-637X/704/2/1274},
  \href {https://ui.adsabs.harvard.edu/abs/2009ApJ...704.1274W} {704, 1274}

\bibitem[\protect\citeauthoryear{{Wolf}, {Martinez}, {Bullock}, {Kaplinghat},
  {Geha}, {Mu{\~n}oz}, {Simon}  \& {Avedo}}{{Wolf} et~al.}{2010}]{Wolf2010}
{Wolf} J.,  {Martinez} G.~D.,  {Bullock} J.~S.,  {Kaplinghat} M.,  {Geha} M.,
  {Mu{\~n}oz} R.~R.,  {Simon} J.~D.,   {Avedo} F.~F.,  2010, \mn@doi [\mnras]
  {10.1111/j.1365-2966.2010.16753.x}, \href
  {https://ui.adsabs.harvard.edu/abs/2010MNRAS.406.1220W} {406, 1220}

\bibitem[\protect\citeauthoryear{{Woo}, {Courteau}  \& {Dekel}}{{Woo}
  et~al.}{2008}]{Woo2008}
{Woo} J.,  {Courteau} S.,   {Dekel} A.,  2008, \mn@doi [\mnras]
  {10.1111/j.1365-2966.2008.13770.x}, \href
  {https://ui.adsabs.harvard.edu/abs/2008MNRAS.390.1453W} {390, 1453}

\makeatother
\end{thebibliography}
%%%%%%%%%%%%%%%%%%%%%%%%%%%%%%%%%%%%%%%%%%%%%%%%%%
%%%%%%%%%%%%%%%%% APPENDICES %%%%%%%%%%%%%%%%%%%%%
\appendix
\section{Orbital Paths}
This appendix presents the orbital paths of our Crater II halo on orbits 1, 2, and 3 as described in Sec.~\ref{sec:orbits}.
\begin{figure}
	\includegraphics[width=1\columnwidth]{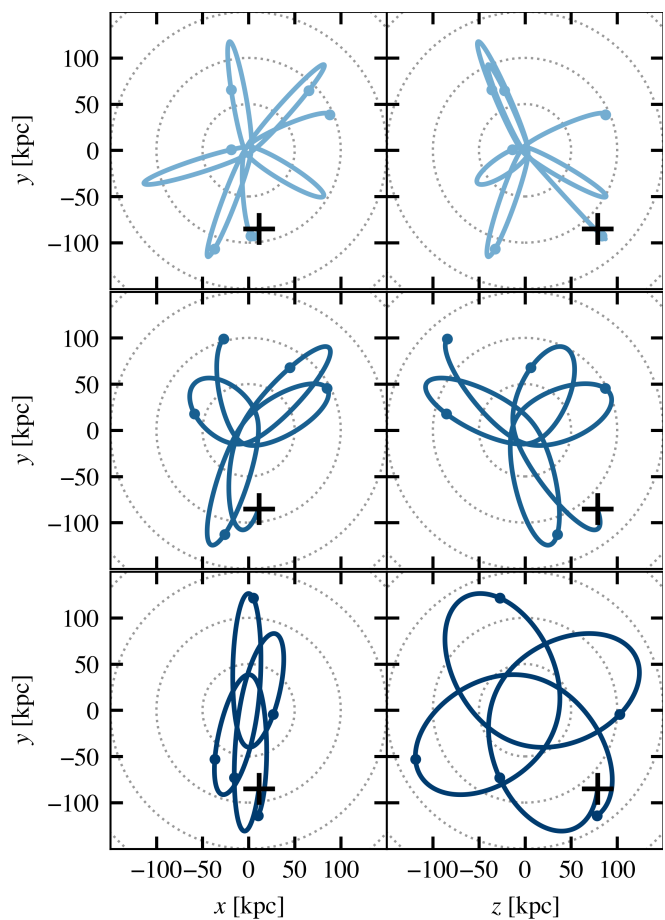}
        \caption{Projected trace on the $X,Y$ (left column) and $Z,Y$ (right column) planes of the three orbits considered in this study (see Table~\ref{tab:CraterII_orbit}; the top row shows orbit 1, middle row -- orbit 2, and bottom row -- orbit 3). The current position of Crater II is indicated by a black cross. Intervals of 2\,Gyr along the orbit are shown using filled circles. The Sun is at $(-8.3,0,0)$ in this coordinate system, with the velocity of the local standard of rest is in the positive $Y$ direction.}
    \label{fig:xyz}
\end{figure}
\section{Supplementary Tidal Tracks}
This appendix presents the supplementary tidal tracks referenced in sections \ref{sec:stars} and \ref{sec:comp_work}.
\begin{figure}
	\includegraphics[width=1\columnwidth]{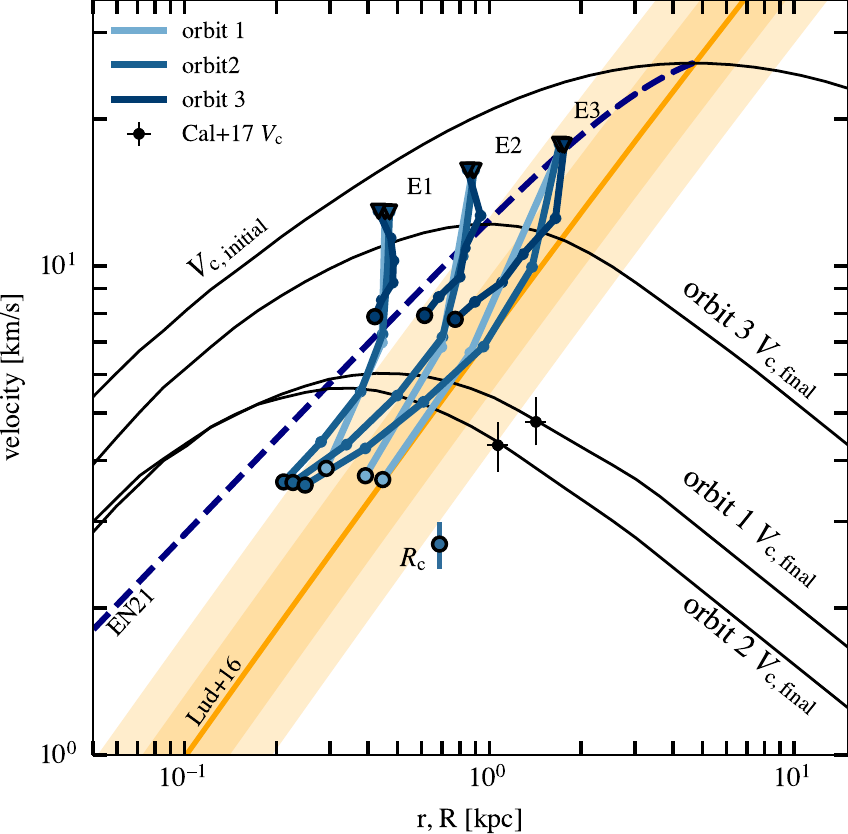}
       \caption{Like Fig.~\ref{fig:tracks}, but with the addition of stellar tracers (E1, E2, and E3 defined in section ~\ref{sec:stars}) tracked through orbits 1 and 3 for comparison.}
    \label{fig:tracks-orbits123}
\end{figure}
\begin{figure}
	\includegraphics[width=1\columnwidth]{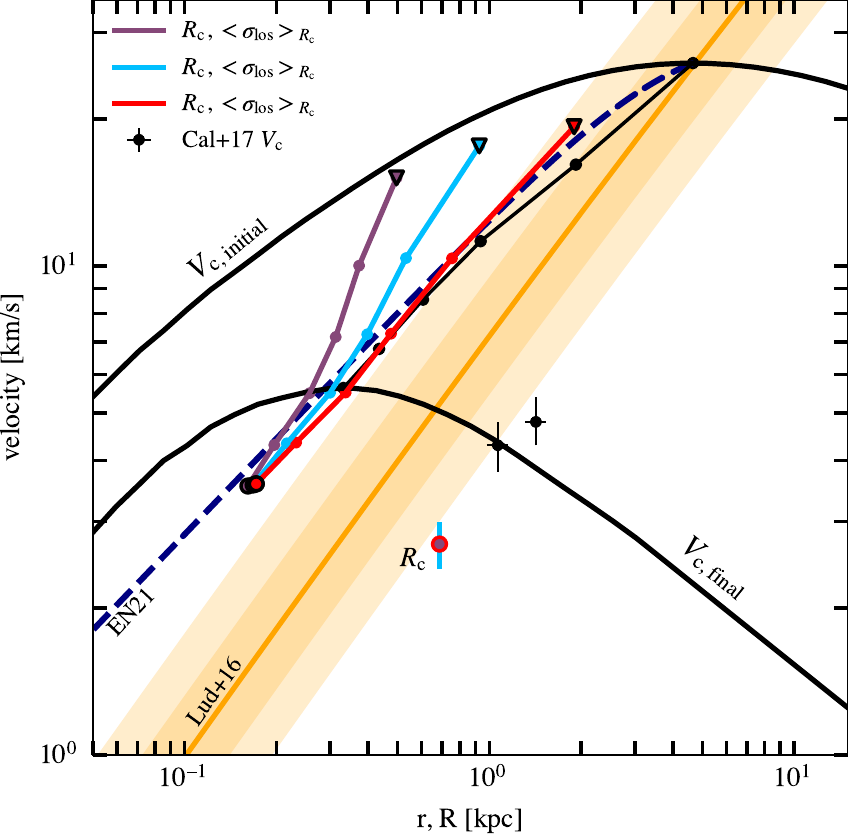}
        \caption{Like Fig.~\ref{fig:tracks}, but for stellar tracers which, at $t=0$, have Einasto $\alpha=0.5$ density profiles. The initial stellar core radii $R_{\rm c}$ are chosen to match those of the exponential tracers in Fig.~\ref{fig:tracks}. Also here, we observe that, in the limit of large tidal mass losses, the stellar core radii and velocity dispersions evolve in sync with the characteristic size and velocity of the underlying DM halo.}
    \label{fig:tracks-alpha0p5}
\end{figure}

%%%%%%%%%%%%%%%%%%%%%%%%%%%%%%%%%%%%%%%%%%%%%%%%%%
% Don't change these lines
\bsp	% typesetting comment
\label{lastpage}
\end{document}